%% file: main.tex
\documentclass[aip,jap,reprint]{revtex4-1}

\usepackage{soul}
\usepackage{bm}
\usepackage{xcolor}
\usepackage{graphicx}
\usepackage{amsmath}
\usepackage{booktabs}
\usepackage{appendix}
\usepackage{multirow}
\usepackage[english]{babel}
\usepackage{multirow}
\usepackage{colortbl}
\usepackage{float}
\usepackage[colorinlistoftodos,textsize=scriptsize]{todonotes}

\begin{document}

\title{Revealing Detailed Transport Mechanisms in Silicon Tunneling Diodes}

\author{Juan P. Mendez}
\email[]{jpmende@sandia.gov}
\affiliation{Sandia National Laboratories, 1515 Eubank Blvd. SE, Albuquerque, NM 87123, USA}
\author{Xujiao Gao}
\email[]{xngao@sandia.gov}
\affiliation{Sandia National Laboratories, 1515 Eubank Blvd. SE, Albuquerque, NM 87123, USA}
\author{Sungjae Hong}
\affiliation{Sandia National Laboratories, 1515 Eubank Blvd. SE, Albuquerque, NM 87123, USA}
\author{Evan M. Anderson}
\affiliation{Sandia National Laboratories, 1515 Eubank Blvd. SE, Albuquerque, NM 87123, USA}
\author{Christopher R. Allemang}
\affiliation{Sandia National Laboratories, 1515 Eubank Blvd. SE, Albuquerque, NM 87123, USA}
\author{Shashank Misra}
\affiliation{Sandia National Laboratories, 1515 Eubank Blvd. SE, Albuquerque, NM 87123, USA}

\date{\today}

\begin{abstract}

Understanding carrier transport in heavily doped semiconductor junctions is essential for developing next-generation tunneling devices for cryogenic electronics and steep-slope technologies. In this work, we investigate the transport mechanisms in a silicon tunneling diode using extensive TCAD (Technology Computer Aided Design) simulations benchmarked against electrical measurements over a wide temperature range of 4~K to 440~K. The measured device characteristics show no negative differential conductance, strong temperature dependence of forward current with a nearly constant semi-logarithmic slope, and weak temperature dependence under high reverse bias. To identify the mechanisms underlying these characteristics, we investigate the contributions of band-to-band tunneling (BTBT), trap-assisted tunneling (TAT), Shockley-Read-Hall (SRH) recombination, and avalanche generation. We also enhance the Schenk TAT model by incorporating Fermi–Dirac statistics to accurately capture the transport behavior at lower temperatures. Under forward bias, TAT in the depletion region dominates the current at low-to-moderate voltages, whereas SRH becomes increasingly significant at higher voltages and temperatures. Under reverse bias, TAT dominates at low reverse voltages, particularly near and above room temperatures, whereas BTBT becomes predominant at high reverse bias, resulting in weak temperature dependence. This work provides a comprehensive understanding of the bias- and temperature-dependent transport mechanisms in silicon tunneling junctions and an enhanced TAT model suitable over a broader range of operating conditions. 

\end{abstract}


\maketitle 

\section{Introduction}\label{sec:introduction}

Heavily doped silicon pn junctions are fundamental building blocks of modern electronics and play a critical role in developing tunneling devices for low-power and steep-slope technologies. \cite{Allemang:2025} Modeling and simulating the characteristics of these junctions using physics-based TCAD (Technology Computer Aided Design) tools is essential for understanding the transport mechanisms underlying the measured device response. Such modeling can help identify limitations in existing TCAD physics models, guide their improvement for more accurate simulations, and ultimately facilitate better device design. In this work, we improve an existing trap-assisted-tunneling model, conduct extensive TCAD simulations, and reveal, over a wide temperature and voltage range, the detailed transport mechanisms governing the device characteristics of silicon diodes fabricated from wafers grown by chemical vapor deposition (CVD). 

Despite the relatively simple geometry of a pn diode, multiple carrier-transport mechanisms, including drift-diffusion (DD), band-to-band tunneling (BTBT), trap-assisted tunneling (TAT), Shockley–Read–Hall (SRH) recombination, and others, can coexist and compete, collectively determining the observed device characteristics.\cite{Hurkx:1992} To distinguish among these competing mechanisms, current–voltage (I–V) characteristics are commonly measured over a broad range of applied voltages and temperatures. For applications that exploit the tunneling current of a diode, considerable effort is devoted to enhancing the BTBT contribution while suppressing competing transport mechanisms. However, achieving a dominant BTBT current can be challenging, particularly under forward-bias conditions, due to limitations associated with device fabrication and/or geometrical design.

\begin{figure}[t]
    \centering
    \includegraphics[width=\linewidth]{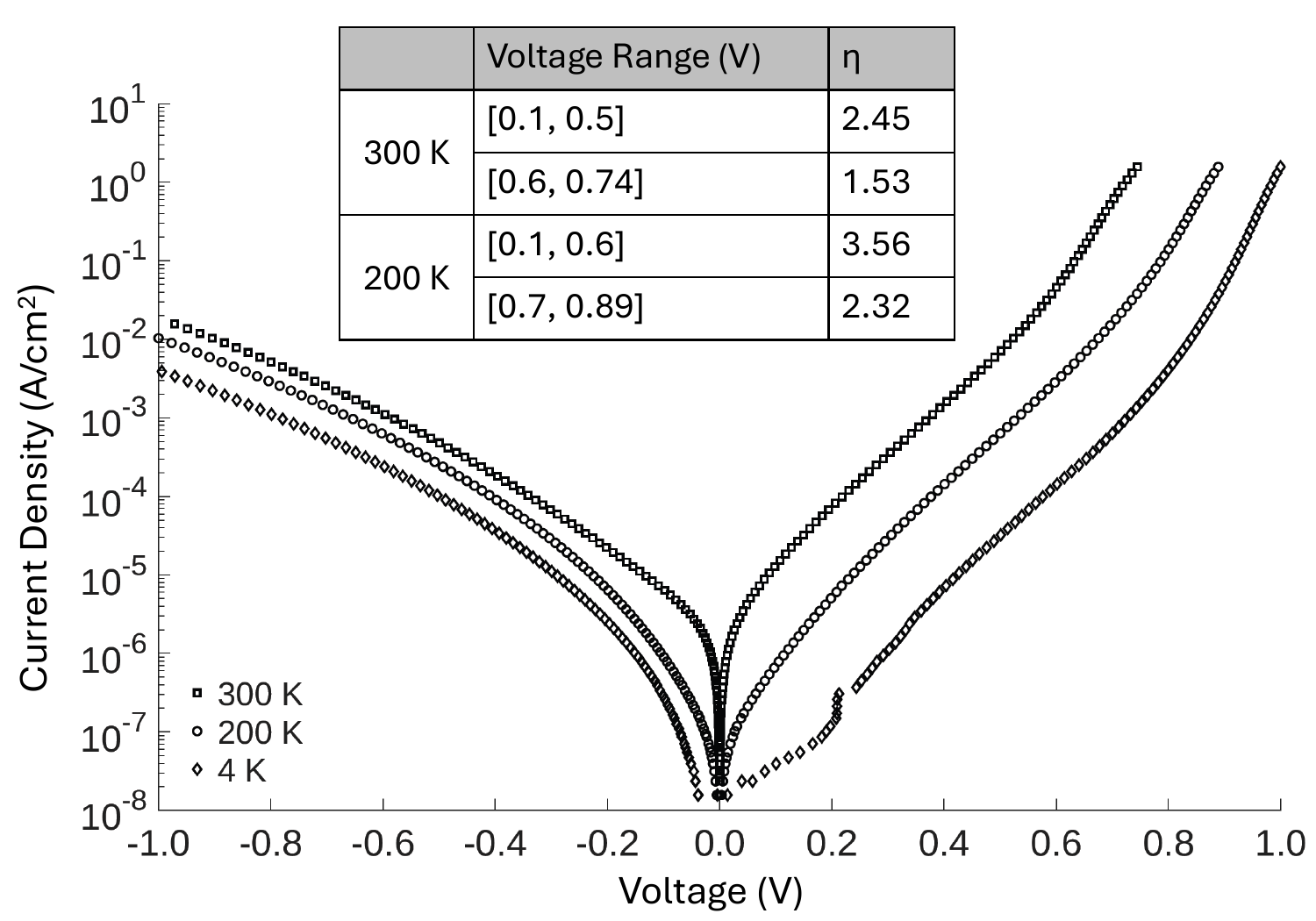}
    \caption{Measured current-voltage characteristics of a CVD-grown silicon pn diode with a diameter of 900 $\mu$m at three temperatures.  The extracted non-ideality factors are shown in the inset and are defined through the diode current-voltage relation, $I = I_0 \textrm{exp} \left( \frac{qV_a}{\eta k_B T} \right)$, where $q$ is the elemental charge, $V_a$ is the applied forward voltage, $k_B$ is the Boltzmann constant, $T$ is the lattice temperature, and $\eta$ is a non-ideality factor, which accounts for deviations from ideal diode behavior and satisfies $\eta > 1$ under non-ideal conditions. }
    \label{fig:Measured-IV-with-eta}
\end{figure}

Fig.~\ref{fig:Measured-IV-with-eta} presents the measured current–voltage (I–V) characteristics of a CVD-grown silicon pn diode under both forward- and reverse-bias conditions at three representative temperatures. Details of the fabrication process for this diode and other pn devices are provided in Ref.~\onlinecite{Hong:2026}. Under forward bias, the absence of a negative differential conductance (NDC) feature indicates that the BTBT contribution is strongly suppressed. In contrast, under reverse bias, the measured current substantially exceeds the expected leakage current of an ideal diode, suggesting a possible contribution from BTBT. Moreover, the I–V characteristics exhibit multiple voltage- and temperature-dependent slopes, providing clear evidence that multiple transport mechanisms contribute to the measured currents. 

We note that similar I-V curves for silicon pn junctions were reported in previous works, \cite{Hurkx:1989,Hurkx:1992} where the authors identified that the measured I-V characteristics were governed by an interplay of TAT, SRH and BTBT mechanisms. In this work, we have not only confirmed the key findings of Hurkx et al., but, more importantly, we have gone beyond their work to discover new findings. On one hand, similar to  Hurkx's findings, our systematical investigation of the different transport mechanisms (i.e., DD, BTBT, TAT, SRH) confirms (i) the measured I-V characteristics in the low-to-medium forward-voltage range (e.g., 0.1 to 0.5 V at 300 K) are determined by the TAT mechanism, which leads to similar slopes in the log(I)-V curves across different temperatures; (ii) SRH recombination dominates the currents at high forward bias (e.g., 0.6 to 0.74 V at 300 K), and its contribution decreases dramatically with decreasing temperatures; and (iii) at room temperature and above, the current is dominated by BTBT under high reverse bias and by TAT under low reverse bias. 

On the other hand, beyond previous work, we have made three additional contributions: (i) we have reproduced the main features of the temperature-dependent experimental data and elucidated the interplay among different transport mechanisms over a wider temperature range, from 4~K to 440~K; (ii) we have identified the applicable temperature limit of the original Schenk TAT model \cite{Schenk:1992} and improved the model to work properly at lower temperatures down to 120 K; (iii) at temperatures of 200 K and below, we have found that reverse-bias I-V characteristics are determined by BTBT, while the contribution of TAT quickly vanishes with decreasing temperatures. This temperature- and voltage-dependent interplay between TAT and BTBT mechanisms produces distinct reverse-bias I-V curves, whereas other mechanisms such as avalanche breakdown have a negligible impact on the reverse-bias currents. 

The paper is organized as follows. In Sec.~\ref{sec: simulation approach}, we describe our TCAD simulation approach with key material models and parameters and explain in detail the improvement of Schenk TAT model for temperatures below 200 K. In Sec.~\ref{sec: transport mechanism}, we investigate each of the important transport mechanisms (i.e., BTBT, TAT, and SRH) under forward-bias conditions, study the effects of their model parameters on the I-V characteristics, and calibrate model parameters to reproduce the measured forward-bias I-V data. In Sec.~\ref{subsection: Reverse response}, we examine how the three mechanisms described above and avalanche breakdown influence the reverse I-V characteristics. Then in Sec.~\ref{sec:temp-dep-transport}, we apply the improved TAT model to simulate the temperature-dependent I-V curves across a wide temperature range and investigate how the dominant transport mechanisms evolve with temperature. Finally, we conclude our work in Sec.~\ref{sec: Conclusions}.

\section{Simulation Approach}\label{sec: simulation approach}

\begin{figure}[t]
    \centering
    \includegraphics[width=1.0\linewidth]{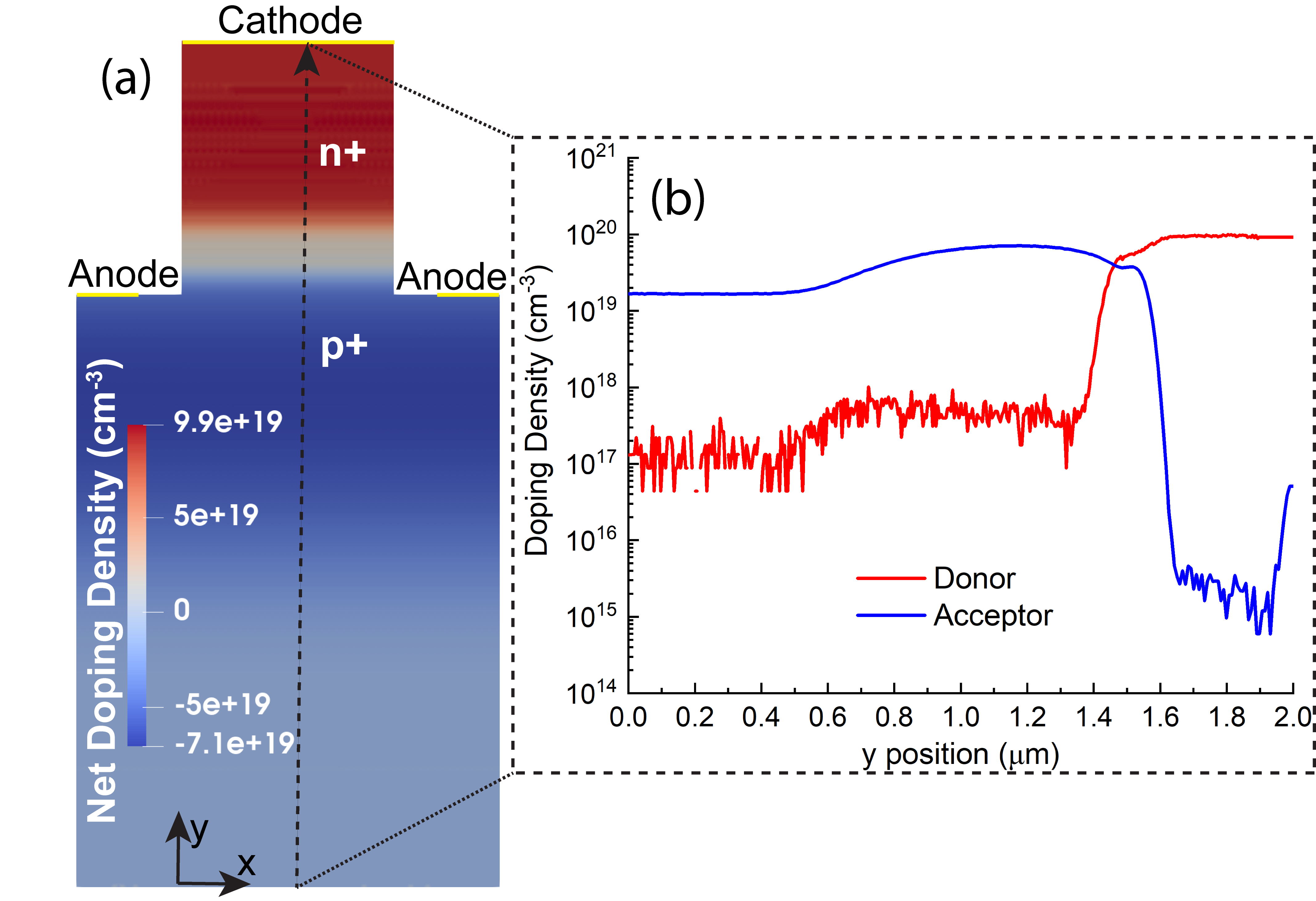}
    \caption{(a) Cross-sectional structure used for TCAD simulations. (b) Measured doping profiles along the line shown in (a).}
    \label{fig:simulation-structure}
\end{figure}

The pn diode analyzed in this work was fabricated from a CVD-grown p$^+$n$^+$ junction on a lightly doped p-type substrate using standard photolithographic processes. From the experimental description in Ref.~\onlinecite{Hong:2026}, we constructed a cross-sectional structure shown in Fig.~\ref{fig:simulation-structure} as a representation of the pn diode, for TCAD simulations. The measured doping profiles in Fig.~\ref{fig:simulation-structure}(b) were obtained through secondary ion mass spectrometry (SIMS) and spreading resistance probe (SRP) measurements, which are necessary inputs for TCAD simulation. To simulate I-V curves, the voltage is applied to the anode contact, while the cathode contact is grounded. Also note that, in our nomenclature, positive bias applied to the anode corresponds to forward bias, whereas negative bias applied to the anode corresponds to reverse bias.

In this work, to investigate the transport mechanism of the fabricated device, we use the open-source TCAD code, Charon, \cite{charon-website,Bochev:2013,Bochev:2013b,Gao:2015,Gao:2026} which is a multi-dimensional, MPI (Message Passing Interface)-parallel, semi-classical device simulation code developed at Sandia National Laboratories. The device geometries and meshes are generated using the tool called Cubit. \cite{Cubit} The simulations are visualized using the open-source data analysis and visualization tool called Paraview. \cite{Paraview}

For our studies, we use the Kane BTBT model, \cite{Kane:1960} the Schenk TAT model\cite{Schenk:1992} and an extended model using Fermi-Dirac (FD)
statistics for low temperatures (described in Sec.~\ref{sec: extended TAT}), the Slotboom model for intrinsic density and band gap narrowing,  \cite{Slotboom:1977} the mid-gap SRH recombination model, the Auger recombination, and the Selberherr avalanche model. \cite{Selberherr:1984,SCHUTZ:1982} For the BTBT model, the parameters are computed according to Ref.~\onlinecite{Kao:2012} for silicon. Additionally, the proposed low-temperature TAT model extends the applicability to lower temperatures than the original Schenk TAT model as described in Sec.~\ref{sec: extended TAT}. For carrier mobility, we use the Arora model \cite{Arora:1982} that takes into account doping dependency. Further details of all other models can be found in the Appendices of this manuscript and in Charon manual\cite{charon-website} for completeness.

\subsection{Enhanced trap-assisted tunneling model}\label{sec: extended TAT}

Two types of TAT models are widely used in TCAD device simulations, including the Hurkx model \cite{Hurkx:1992} and the Schenk model. \cite{Schenk:1992} Both models use an SRH-like recombination expression and field-dependent lifetimes to capture the effect of TAT. The SRH-like expression for TAT is given as
\begin{equation}
R_{TAT} = \sum_j \frac{np-n_{ie}^2}{\tau_p^j(n+n_t^j)+\tau_n^j(p+p_t^j)}, 
\label{eq:tat-eq-1}
\end{equation}
where the summation runs over the total number of different types of traps. $n$ is the electron density, $p$ is the hole density, $n_{ie}$ is the effective intrinsic carrier concentration of the material, $n_t^j$ and $p_t^j$ for the j-th type of trap are equal to
\begin{equation}
n_t^j=N_C \exp{\bigg( -\frac{E_t^j}{k_BT} \bigg)},~p_t^j=N_V\exp{\bigg( -\frac{E_g-E_t^j}{k_BT} \bigg)},
\label{eq:tat-eq-2}
\end{equation}
where $N_C$ and $N_V$ are the effective density of states in the conduction and valence bands, respectively, $E_t^j$ is the j-th trap energy measured from the conduction band edge, and $E_g$ is the effective band gap. The lifetimes $\tau_n^j$ and $\tau_p^j$ depend on the lattice temperature and the electric field
\begin{equation}
\tau_n^j(T,F)=\frac{\tau_{n0}^j \left(\frac{T}{300} \right)^\alpha}{1+g_n^j(T,F)},~\tau_p^j(T,F)=\frac{\tau_{p0}^j \left(\frac{T}{300
} \right)^\alpha}{1+g_p^j(T,F)},
\label{eq:tat-eq-3}
\end{equation}
where $\tau_{n0}^j$ and $\tau_{p0}^j$ are field- and temperature-independent lifetimes with $\alpha=-1.5$ for Si. $g_n^j(T,F)$ and $g_p^j(T,F)$ are the electron and hole field enhancement factors, which capture the recombination enhancement due to band-to-trap tunneling. Note that Eq.~(\ref{eq:tat-eq-1}) is obtained assuming the Maxwell-Boltzmann (MB) statistics. Both Hurkx and original Schenk TAT models use MB statistics to derive the field enhancement factors.

Within the Hurkx TAT model, \cite{Hurkx:1992} the field enhancement factors are computed without any phonon information, which means the TAT effect is modeled without phonon assistance. This is valid for direct semiconductors such as gallium arsenide. However, for indirect semiconductors such as silicon, a multiphonon TAT process is more plausible, \cite{Schenk:1992} hence we adopted the Schenk TAT model for the silicon device studied in this work.   

The electron field enhancement factor from the Schenk model with the MB statistics takes the form of \cite{Gao:2019}
\begin{equation}
\label{eq:gn-MB}
g_n(T,F) = \frac{\int_0^{E_t}  \rho_n(E) I_{E/\hbar\omega_0}(z) \exp \left(\frac{- E} {2 k_B T} \right) dE} {\int_{E_t}^\infty \rho_n^{F=0}(E) I_{E/\hbar\omega_0}(z) \exp \left(\frac{- E} {2 k_B T} \right) dE}.
\end{equation}
For a simpler notation, the superscript $j$ is dropped from here on. $I_{E/\hbar\omega_0}(z)$ is 
the modified Bessel function. For large orders (i.e., $l = E/\hbar
\omega_0 \gg 1$, which generally holds for deep-level traps that are
relevant for band-to-trap tunneling), it can be approximated by
\begin{eqnarray} \label{eq:Il-definition}
I_l(z) &=& \frac{1}{\sqrt{2 \pi}} \frac{1}{(l^2+z^2)^{1/4}} \exp \left(\sqrt{l^2+z^2} \right) \nonumber \\
&& \exp \left[ -l \ln \left( \frac{l}{z} + \sqrt{1 + \frac{l^2}{z^2}}  \right) \right],
\end{eqnarray}
with
\begin{equation} \label{eq:zandfB-definition}
z =  2S \sqrt{f_B(f_B+1)}, \; \; f_B = \left[ \exp \left( \frac{\hbar \omega_0}{k_B T} \right) -1 \right]^{-1}.
\end{equation}
$f_B$ is the Bose distribution function. The zero-field density of
states (DOS), $\rho_n^{F=0}(E)$, is given by
\begin{equation}
\label{eq:zerofieldLDOS}
\rho_n^{F=0}(E) =  \frac{1}{2 \pi^2} \left( \frac{2 m_n^*}{\hbar^2} \right)^{3/2} \sqrt{E - E_t}.
\end{equation}
The field-dependent density of states, $\rho_n(E)$, in the Schenk model was derived based on the constant field assumption and is equal to
\begin{eqnarray}
\label{eq:constfdos}
\rho_n(E) &=& \frac{1}{2 \pi} \left( \frac{2 m_n^*}{\hbar^2} \right)^{3/2} \sqrt{\hbar \Theta} \mathcal{F}\left( \frac{E_t-E} {\hbar \Theta} \right), \nonumber \\ 
 \hbar \Theta &=& \left(\frac{\hbar^2 q^2 F^2} {2 m_n^*} \right)^{1/3}, \nonumber \\
 \mathcal{F}(y) &=& A_i^{\prime 2}(y) - y A_i^2(y).
\end{eqnarray}
$F$ is the electric field magnitude, while $\mathcal{F}(y)$ is known as the electrooptical function. $A_i(\cdot)$
is the Airy function of the first kind. $A_i^\prime(\cdot)$ is the
first derivative of the Airy function. 
Analogous results for $g_p(T, F)$ can be obtained by using the
hole DOS and $(E_g - E_t)$.

As shown in Sec.~\ref{subsec:TAT}, TCAD simulations using the above MB-based Schenk TAT model produce well-behaved current-voltage results. And in Sec.~\ref{subsec:comparison}, we show that simulation results obtained using the same set of TAT parameters can match the measured data very well at both 300 K and 200 K. Clearly, the Schenk TAT model with MB statistics works well at these relatively high temperatures. However, when the temperatures become smaller than 200 K, the use of MB statistics in the TAT model produces nonphysical results as discussed in Appendix \ref{Appendix:Stat-Effect-in-TAT}. Hence, we propose to improve the original Schenk TAT model by including the Fermi-Dirac (FD) statistics. In the following, we provide a detailed description of the proposed extension.

First, the TAT recombination rate expression needs to include the FD statistics and is given by \cite{Schenk:1992}
\begin{equation}
\label{eq:Rtat-FD}
R_{TAT} = \frac{np \left[ 1 - \textrm{exp}\left(-\frac{E_{Fn}-E_{Fp}}{k_BT}  \right) \right]} {\tau_{p}(T,F) \frac{n}{f_{tn}} + \tau_{n}(T,F) \frac{p}{(1-f_{tp})} }.
\end{equation}
For clarity, we consider a single trap level and omit the trap index $j$. Here $E_{Fn}$ ($E_{Fp}$) is the electron (hole) quasi-Fermi energy level, $f_{tn}$ ($f_{tp}$) is the trap occupation probability for electrons (holes) and is defined as,
\begin{equation}
f_{t \nu} = \left[ 1 + \textrm{exp} \left( \frac{E_C - E_t - E_{F \nu}} {k_B T} \right) \right]^{-1}, \; \; \nu = n, p
\end{equation}
with $E_C$ being the conduction-band edge. 
In Charon TCAD code, the electron density $n$ and hole density $p$ are directly solved for, while the quasi-Fermi energy levels, $E_{F \nu}$ ($\nu = n, p$), are related to the carrier densities. Therefore, we rewrite all the quantities involving quasi-Fermi levels in Eq.~(\ref{eq:Rtat-FD}) in terms of carrier densities and other material properties as shown below  
\begin{eqnarray}
  &1 - \textrm{exp}\left(-\frac{E_{Fn}-E_{Fp}}{k_BT}  \right)= \nonumber \\
  &1 - \textrm{exp} \left[-\frac{E_g}{k_BT} - \mathcal{F}^{-1}_{1/2} \left( \frac{n}{N_C}\right)  - \mathcal{F}^{-1}_{1/2} \left( \frac{p}{N_V}\right) \right]
\end{eqnarray}

\begin{eqnarray}
    f_{tn} = \left[1 + \textrm{exp}\left(-\mathcal{F}^{-1}_{1/2} \left( \frac{n}{N_C}\right) \right) \cdot \textrm{exp} \left(-\frac{E_t}{k_BT} \right)  \right]^{-1}
\end{eqnarray}

\begin{eqnarray}
    &1- f_{tp} = \\
    &\left[1 + \textrm{exp}\left(-\mathcal{F}^{-1}_{1/2} \left( \frac{p}{N_V}\right) \right) \cdot \textrm{exp} \left(-\frac{E_g - E_t}{k_B T} \right)  \right]^{-1},
\end{eqnarray}
using the following relations
\begin{align}
\mathcal{F}_{1/2}^{-1}\left(\frac{n}{N_C}\right)
=
\frac{E_{Fn}-E_C}{k_BT},\\
\qquad
\mathcal{F}_{1/2}^{-1}\left(\frac{p}{N_V}\right)
=
\frac{E_V-E_{Fp}}{k_BT}.
\end{align}

Second, we need to include the FD statistics in the calculation of field enhancement factors. Note the formula in Eq.~(\ref{eq:gn-MB}) for the enhancement factors was obtained using the MB statistics. To include the FD statistics, we recall the original expressions \cite{Gao:2019}
\begin{eqnarray}
\label{eq:gn-general}
g_n(T,F) &=& \frac{\int_0^{E_t}  \rho_n(E) M_n(E) f_n(E) dE} {\int_{E_t}^\infty \rho_n^{F=0}(E)  M_n(E) f_n(E) dE}, \\
g_p(T,F) &=& \frac{\int_0^{E_g - E_t}  \rho_p(E) M_p(E) (1-f_p(E)) dE} {\int_{E_g - E_t}^\infty \rho_p^{F=0}(E)  M_p(E) (1 - f_p(E)) dE}. \nonumber \\ 
\end{eqnarray}
Here $M_n$ ($M_p$) is the multiphonon transition probability for electron (hole) capture, and $f_n$ ($f_p$) is the electron (hole) FD distribution function. The multiphonon transition probability can be expressed as \cite{Schenk:1992}
\begin{equation} \label{eq:tat-trans-prob}
    M_\nu(E)  = \textrm{exp}[-S(2 f_B +1)] \textrm{exp} \left( \frac{E} {2 k_B T} \right) I_l(z), \nu=n, p
\end{equation}
where $S$ is the Huang-Rhys factor, which is a measure of the diagonal electron-phonon coupling strength, $I_l(z)$ and $f_B$ are defined in Eq.~(\ref{eq:Il-definition}) and Eq.~(\ref{eq:zandfB-definition}), respectively. The carrier FD distribution functions can be written as 
\begin{eqnarray} \label{eq:tat-elec-fd}
f_n(E) &=& \left[ 1 + \textrm{exp} \left( \frac{E_C - E_t + E - E_{Fn}} {k_B T} \right) \right]^{-1} \nonumber \\
       &=& \left[1 + \textrm{exp}\left(-\mathcal{F}^{-1}_{1/2} \left( \frac{n}{N_C}\right) \right) \cdot \textrm{exp} \left(-\frac{E_t}{k_BT} \right) \right . \nonumber \\
       &\cdot& \left. \textrm{exp} \left(\frac{E}{k_BT} \right) \right]^{-1},
\end{eqnarray}

\begin{eqnarray} \label{eq:tat-hole-fd}
 1- f_p(E) &=& 1 - \left[ 1 + \textrm{exp} \left( \frac{E_C - E_t - E - E_{Fp}} {k_B T} \right) \right]^{-1} \nonumber \\
        &=& \left[1 + \textrm{exp}\left(-\mathcal{F}^{-1}_{1/2} \left( \frac{p}{N_V}\right) \right) \right.  \nonumber \\
        &\cdot& \left. \textrm{exp} \left(-\frac{E_g - E_t}{k_B T} \right) \cdot \textrm{exp} \left(\frac{E}{k_B T} \right) \right]^{-1}.
\end{eqnarray}

\begin{figure}[t]
    \centering
    \includegraphics[width=1\linewidth]{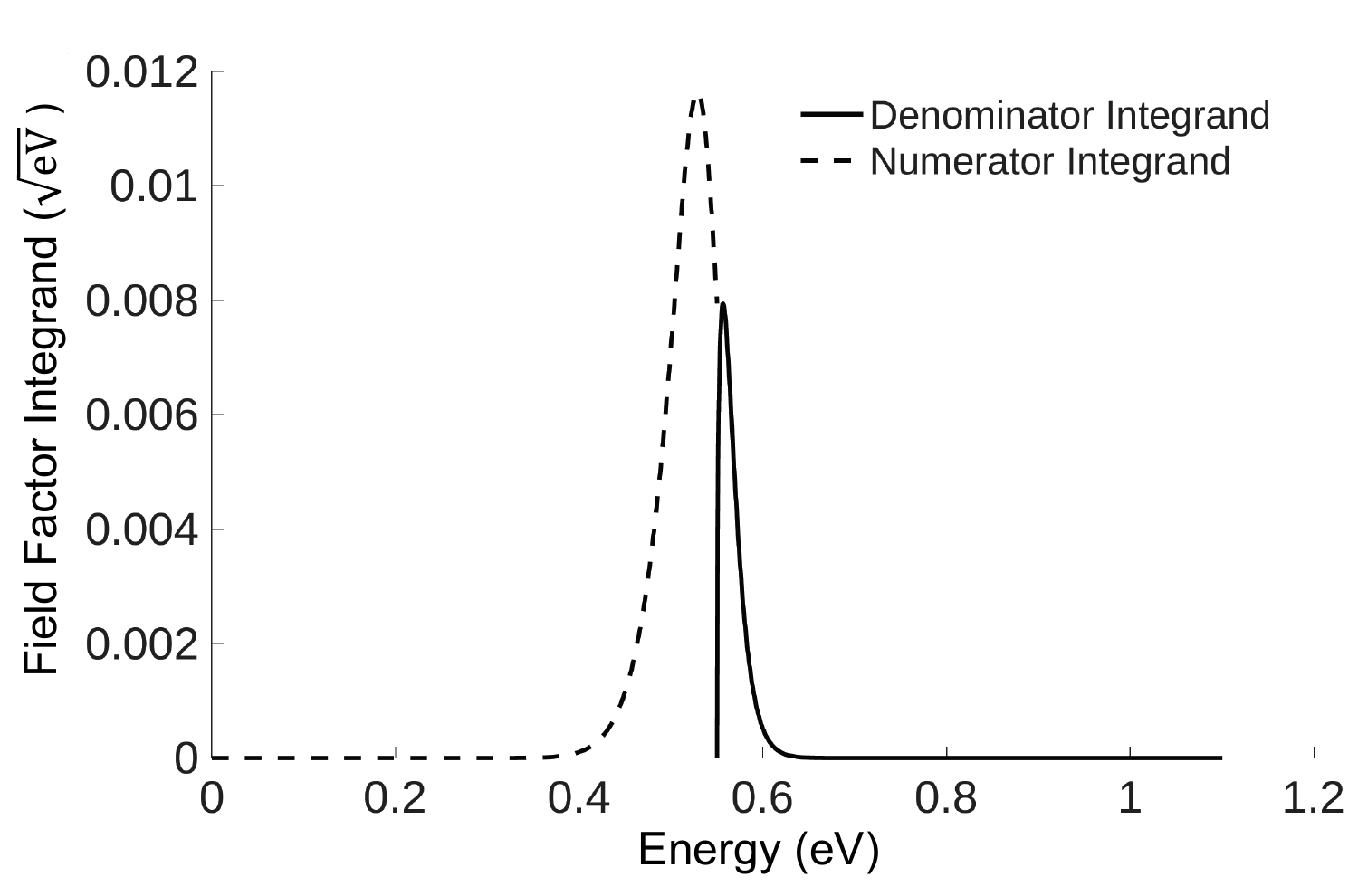}
    \caption{An example showing the energy-dependent denominator and numerator integrands of the electron field enhancement factor. For the example, we arbitrarily chose $E_t = 0.55 $ eV, T = 150 K, $F = 4 \times 10^5$ V/cm, and $n = 10^{19}$ cm$^{-3}$. The numerator integrand goes from 0 to 0.55 eV, while the denominator integrand starts from 0.55 eV to $2 \times 0.55$ eV.}
    \label{fig:ff-integrand}
\end{figure}

Substituting Eqs.~(\ref{eq:tat-trans-prob}), (\ref{eq:tat-elec-fd}), (\ref{eq:tat-hole-fd}), and the DOS expressions in Eqs.~(\ref{eq:zerofieldLDOS}), (\ref{eq:constfdos}) into Eq.~(\ref{eq:gn-general}), we can compute the electron and hole field enhancement factors by performing integration over energy. Note that Charon is an MPI-based (Message Passing Interface) parallel TCAD code, which means a device structure is decomposed into many spatial domains during a simulation run. This type of domain decomposition makes it very challenging to perform integration over space, but integration over energy can be easily done. Fig.~\ref{fig:ff-integrand} shows an example of the energy-dependent denominator and numerator integrands of the electron field enhancement factor in Eq.~(\ref{eq:gn-general}). Clearly, these integrands are smooth functions of energy and the denominator integrand decays exponentially toward zero at high energies, hence the energy upper bound for the denominator integration can be restricted. In our study, we find that $2 E_t$ is a sufficient upper bound for the denominator integration without sacrificing any accuracy. Similarly, an energy upper bound of $2(E_g-E_t)$ was used for the denominator of the hole field enhancement factor. Energy integration can be done using the simple trapezoidal rule. In our work, we used the trapezoidal rule with 200 energy points and noticed negligible increase in simulation time. On the other hand, due to the smoothness of the integrands, an efficient quadrature rule can be explored to speed up large simulations.

The proposed FD-based TAT model was employed to simulate currents in the pn device at 175 K and 150 K, yielding physically reasonable results, as discussed in Appendix \ref{Appendix:Stat-Effect-in-TAT}. However, further investigation indicates that when the temperature falls below 120 K, simulations using either the MB-based or FD-based TAT model become difficult to converge because of the complex exponential temperature dependence. Additional research is needed to improve existing TAT models or develop new ones capable of accurately simulating TAT effects at temperatures below 120 K and into the cryogenic regime. Addressing this challenge will require continued research efforts in the future. In this work, because the MB-based TAT model performs well at temperatures of 200 K and above, it was used for all simulations at these relatively high temperatures. For temperatures ranging from 120 K to 200 K, the FD-based TAT model was employed for forward-bias simulations. 

\section{Transport Mechanism for Forward Bias}\label{sec: transport mechanism}

In this section, we investigate the various carrier transport mechanisms that may contribute to the electrical response of the fabricated diode under forward-bias conditions. In particular, we examine BTBT, TAT, and SRH recombination. These mechanisms have previously been identified as the dominant transport processes in pn junction diodes. \cite{Chynoweth:1961,Hurkx:1989,Hurkx:1992,Jorke:1993,Reitemann:2002,Jandieri:2008,Oehme:2010,Schmid:2012}

Our analysis proceeds as follows. We first evaluate the contributions of the different transport mechanisms at 300~K under forward-bias conditions. Based on these simulations, we identify the dominant transport mechanisms responsible for the measured device responses by comparing the simulated characteristics against the experimental data and calibrating the model parameters accordingly. We then extend the analysis to lower temperatures.

\subsection{Band-to-band tunneling}

\begin{figure}[t]
    \centering
    \includegraphics[width=1\linewidth]{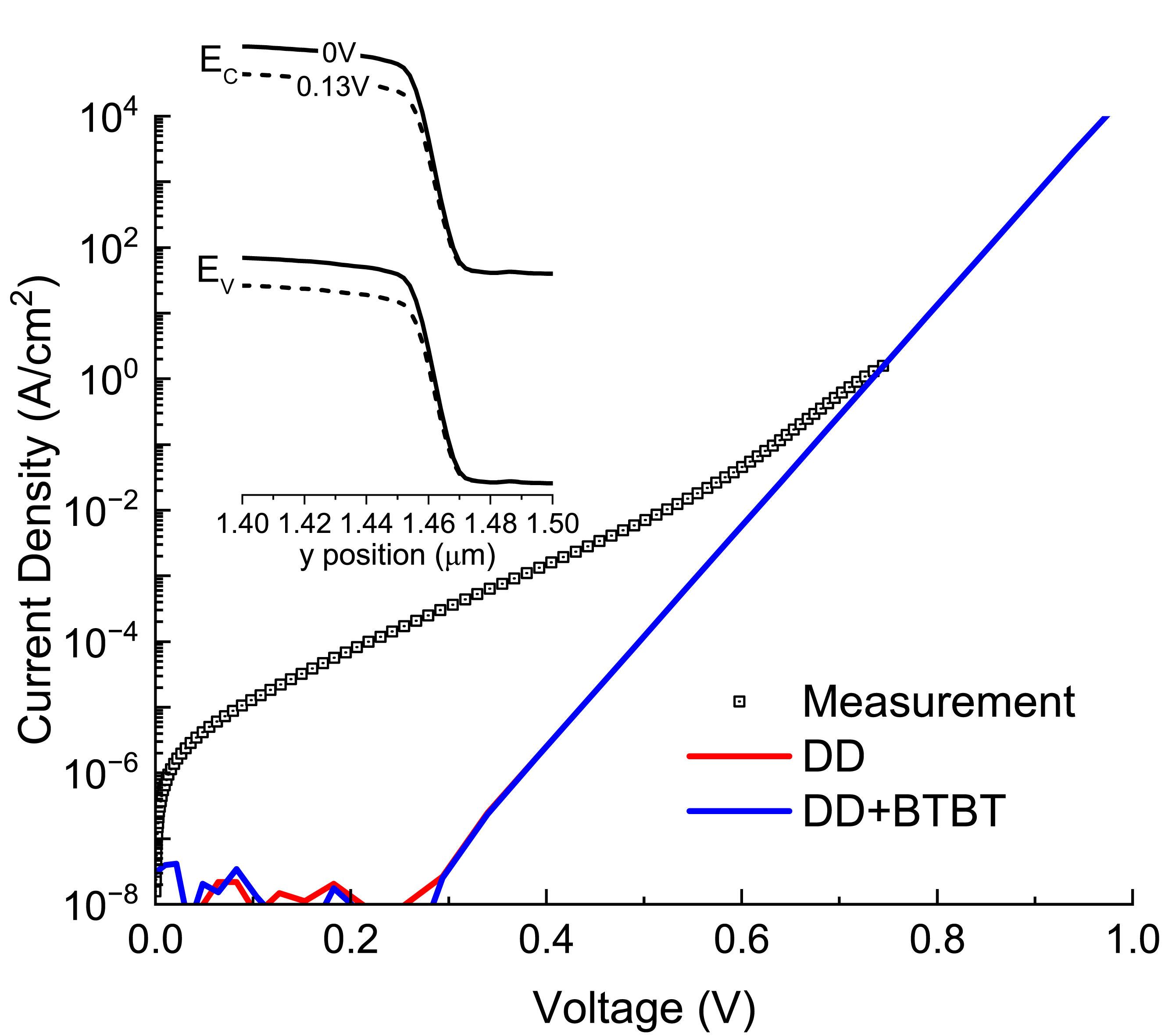}
    \caption{Analysis of the BTBT contribution in the forward-bias regime at T=300~K. Inset: Band  structure across the junction under equilibrium and an applied voltage of 0.13~V. E$_C$ and E$_V$ indicate the conduction- and valence-band edges, respectively.}
    \label{fig:BTBT}
\end{figure}

In highly doped and sufficiently abrupt pn junctions, negative differential conductance (NDC) in the measured forward-bias I-V characteristics can be observed when BTBT dominates the current in the low to medium voltage range. \cite{Oehme:2009,Oehme:2010,Oehme:2012,Schmid:2012} The origin of the NDC behavior is twofold: (i) at sufficiently low forward bias, the BTBT current increases as more occupied conduction-band states on the n-type side become energetically aligned with empty valence-band states on the p-type side with increasing applied bias; and (ii) as the forward bias is increased further, this overlap decreases, reducing the number of available tunneling states and causing the current to decline despite the increasing applied bias, thereby giving rise to the NDC region in the current-voltage characteristics. Therefore, the absence of NDC in the measured forward-bias I-V curve in Fig.~\ref{fig:Measured-IV-with-eta} suggests that BTBT is not the dominant transport mechanism in the present device under forward-bias conditions. 
 
To verify the insignificant contribution of BTBT under forward bias in this device, we performed TCAD simulations using the Kane BTBT model, described in Appendix~\ref{Appendix:BTBT}. In previous work, \cite{Mendez:2025} Charon simulations using the Kane BTBT model have successfully reproduced the measured I-V characteristics and key transport features of atomic precision advanced manufacturing (APAM)-enabled pn junctions. \cite{Skeren:2020} Fig.~\ref{fig:BTBT} compares the simulated BTBT current with the drift-diffusion-only current. The results show that BTBT contribution is indeed negligible over the entire forward-bias range investigated, as the simulated current is essentially identical to the drift-diffusion current.

The weak BTBT contribution can be attributed to the relatively broad junction transition region produced by the CVD growth process (see Fig.~\ref{fig:simulation-structure}). Although the peak doping concentrations exceed 10$^{20}$ cm$^{-3}$, the junction is not sufficiently abrupt as shown by the doping profiles in Fig.~\ref{fig:simulation-structure}(b) to generate the strong electric field and band overlap required for significant interband tunneling. This is further illustrated by the band structure shown in the inset of Fig.~\ref{fig:BTBT}. At equilibrium (i.e., zero applied bias), the spatial separation across the junction between the conduction-band edge (E$_C$) on the n-type side and the valence-band edge (E$_V$) on the p-type side is approximately 20~nm, resulting in a tunneling distance that is too large for significant BTBT tunneling. In addition, the small energetic overlap between the conduction band on the n-type side and the valence band on the p-type side quickly diminishes as the forward bias is increased. Consequently, the device does not exhibit the NDC characteristic of Esaki diodes. This observation is consistent with previous studies showing that BTBT becomes dominant only in highly doped and abrupt pn junctions or p$^+$-i-n$^+$ structures with narrow intrinsic regions, where the tunneling probability is substantially enhanced as narrower intrinsic regions produce stronger electric fields and shorter tunneling distances. \cite{Jorke:1993,Oehme:2009,Oehme:2012}

\subsection{Trap-assisted tunneling}\label{subsec:TAT}

Previous studies on silicon Esaki and p$^+$-i-n$^+$ junctions have demonstrated that TAT is an important transport mechanism at low-to-moderate forward biases and plays a key role in explaining the excess current\cite{Oehme:2009,Oehme:2012} which cannot be reproduced by conventional SRH alone.\cite{Hurkx:1989,Hurkx:1992,Reitemann:2002,Oehme:2012,Schmid:2012}

TAT is a field-enhanced transport mechanism in which charge carriers tunnel through localized defect states within the band gap under strong electric fields, with phonons assisting the process when momentum conservation is required. In the Schenk formalism,\cite{Schenk:1992} the electric-field dependence is incorporated through field enhancement factors that effectively reduce the carrier lifetime  under strong electric fields, and therefore increase the recombination rate. In heavily doped junctions, the large built-in electric field across the junction can significantly enhance the TAT transport, particularly at low and moderate forward biases.\cite{Hurkx:1989,Hurkx:1992,Reitemann:2002,Oehme:2012,Schmid:2012} Previous studies have also shown that TAT strongly depends on both defect density and trap energy level. In addition, TAT transport has often been correlated with the depletion width in pn junctions, with stronger TAT observed for narrower depletion regions.\cite{Oehme:2012,Jorke:1993} 
This correlation arises because narrower depletion regions result in stronger electric fields across the junction, which enhance the tunneling probability and, consequently, the TAT contribution.

\begin{figure}[t]
    \centering
    \includegraphics[width=1\linewidth]{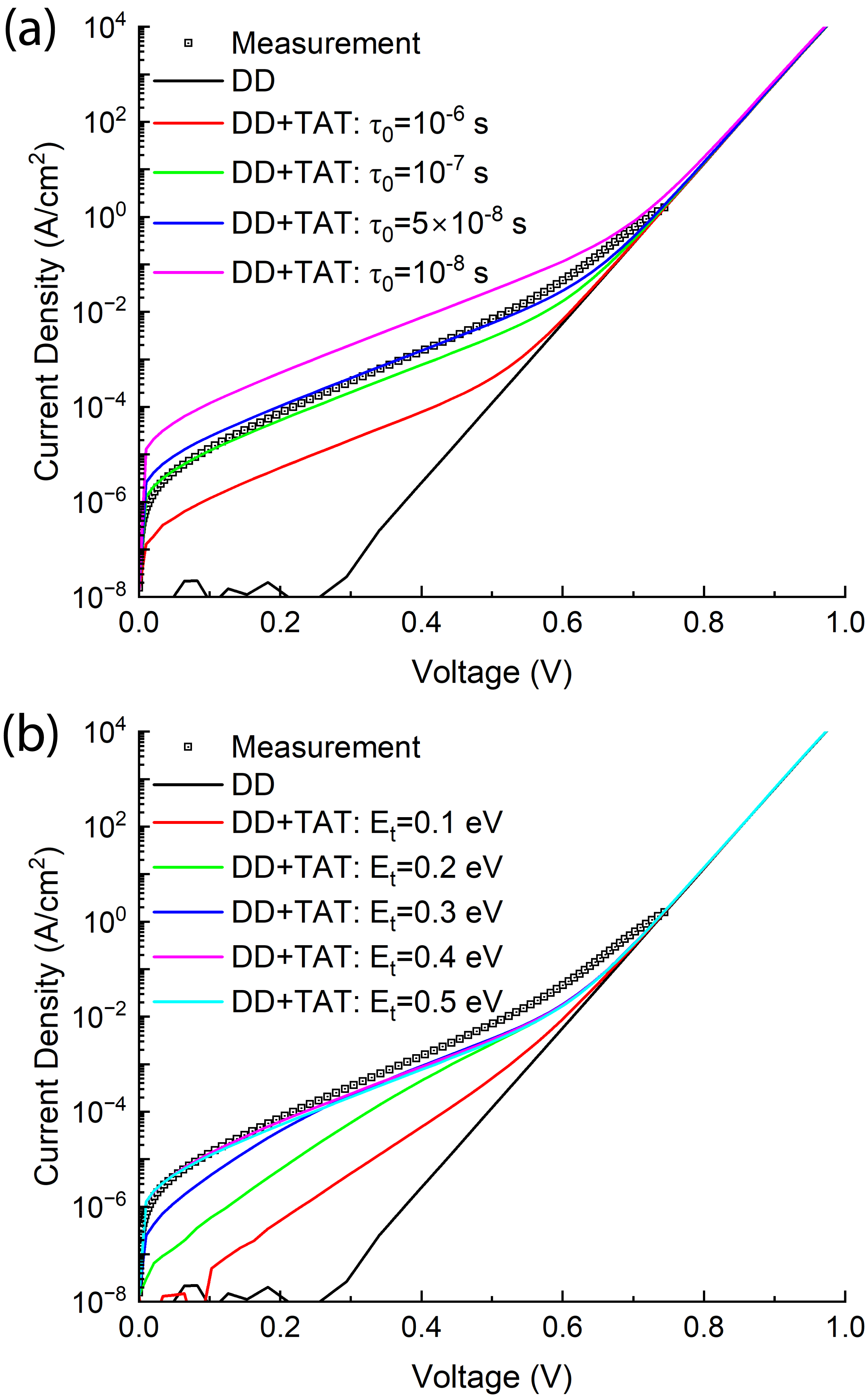}
    \caption{Analysis of the TAT contribution in the forward-bias regime at T=300~K: (a) effect of field-independent lifetime $\tau_0$ for a fixed trap energy level of $E_t=0.5$~eV, and (b) effect of trap energy level $E_t$ for a fixed field-independent lifetime of $\tau_0=10^{-7}$~s. Note that the TAT simulations take into account the actual field-dependent lifetimes following Eq.~(\ref{eq:tat-eq-3}).}
    \label{fig:TAT} 
\end{figure}

Fig.~\ref{fig:TAT} shows the analysis of the field-independent lifetime $\tau_0$ and trap energy level $E_t$ in panels (a) and (b), respectively. For simplicity, we assume $\tau_{n0} \approx \tau_{p0} = \tau_0$. The field-independent lifetime is related to the trap density through $\tau_0 = 1 / (\sigma v N_t)$, where $\sigma$ is the carrier capture cross section, $v$ is the carrier thermal velocity, and $N_t$ is the trap density.
We note that the TAT simulations account for the effective field-dependent lifetimes following Eq.~(\ref{eq:tat-eq-3}). The TAT contribution in Fig.~\ref{fig:TAT} can be identified as the difference between the TAT-plus-drift-diffusion currents and the drift-diffusion-only current.  

In Fig.~\ref{fig:TAT}(a), reducing $\tau_0$ primarily increases the TAT current magnitude while preserving an approximately constant semi-logarithmic current-voltage slope across the entire bias range. Since the field-independent lifetime is inversely proportional to the trap density, a smaller $\tau_0$ corresponds to a higher concentration of defects near the junction, leading to an enhanced trap-assisted recombination rate and, consequently, a larger TAT current. In addition, the approximately constant semi-logarithmic current-voltage slope with varying $\tau_0$ can be understood by noting that, when $\tau_{0n} \approx \tau_{0p}= \tau_{0}$ is assumed in Eq.~(\ref{eq:tat-eq-1}), $1/\tau_0$ acts primarily as a bias-independent multiplier in the TAT recombination rate. Thus, changing $\tau_0$ mainly scales the current magnitude without significantly changing its voltage dependence.  

Fig. \ref{fig:TAT}(b) shows the effect of trap energy $E_t$ (measured from the conduction band edge) on the characteristic curves. Clearly, trap states located near mid-gap contribute most effectively to the TAT current. In particular, the TAT contribution increases as the trap energy level moves deeper into the band gap, reaching a maximum near mid-gap.  We also note, as the trap moves closer to mid-band gap, the effect of $E_t$ on the current is saturated, which is evident by the fact that the log(I)-V curves for $E_t = 0.4$ eV and $E_t = 0.5$ eV are nearly identical. These observations are consistent with the fact that mid-gap states maximize carrier capture probability in SRH processes. \cite{Sze:Chapter1} Unlike the effect of $\tau_0$, the trap energy $E_t$ has a strong influence on the log(I)-V slopes as well as the current values. This is because $E_t$ affects the TAT recombination rate in a complex manner as seen from the mathematical expressions in Sec.~\ref{sec: extended TAT}.

By comparing the TAT simulations (Fig.~\ref{fig:TAT}) with the measurements (Fig.~\ref{fig:Measured-IV-with-eta}), we find that the measured current can be reproduced very well over the low- to moderate-forward-bias regime (approximately 0-0.45~V), but not at higher biases (approximately above 0.45~V). This indicates that additional transport mechanisms become relevant as the electric field across the junction decreases and the tunneling enhancement weakens. In the following, we investigate the SRH transport mechanism over this bias regime.

\subsection{Shockley-Read-Hall recombination}

SRH recombination can become particularly important under low electric-field conditions, where the contributions from BTBT and TAT are weaker. Note that SRH discussed in this work is referred to the recombination effect from mid-band gap defect states since they act as the most efficient recombination centers in zero- and small-field regions. \cite{Sze:Chapter1} Previous studies have shown that SRH recombination dominates at higher forward bias below the diode threshold voltage, where the built-in electric field across the junction is reduced.\cite{Hurkx:1989,Hurkx:1992} Ref.~\onlinecite{Jorke:1993} further reported that the SRH current dominates in p$^+$-i-n$^+$ junctions with wide intrinsic regions, whereas TAT and eventually BTBT become dominant as the intrinsic-region width decreases. Physically, wider intrinsic regions reduce the electric field across the junction and increase the tunneling distance, thereby suppressing tunneling-assisted transport mechanisms such as TAT and BTBT.

\begin{figure}[t]
    \centering
    \includegraphics[width=1\linewidth]{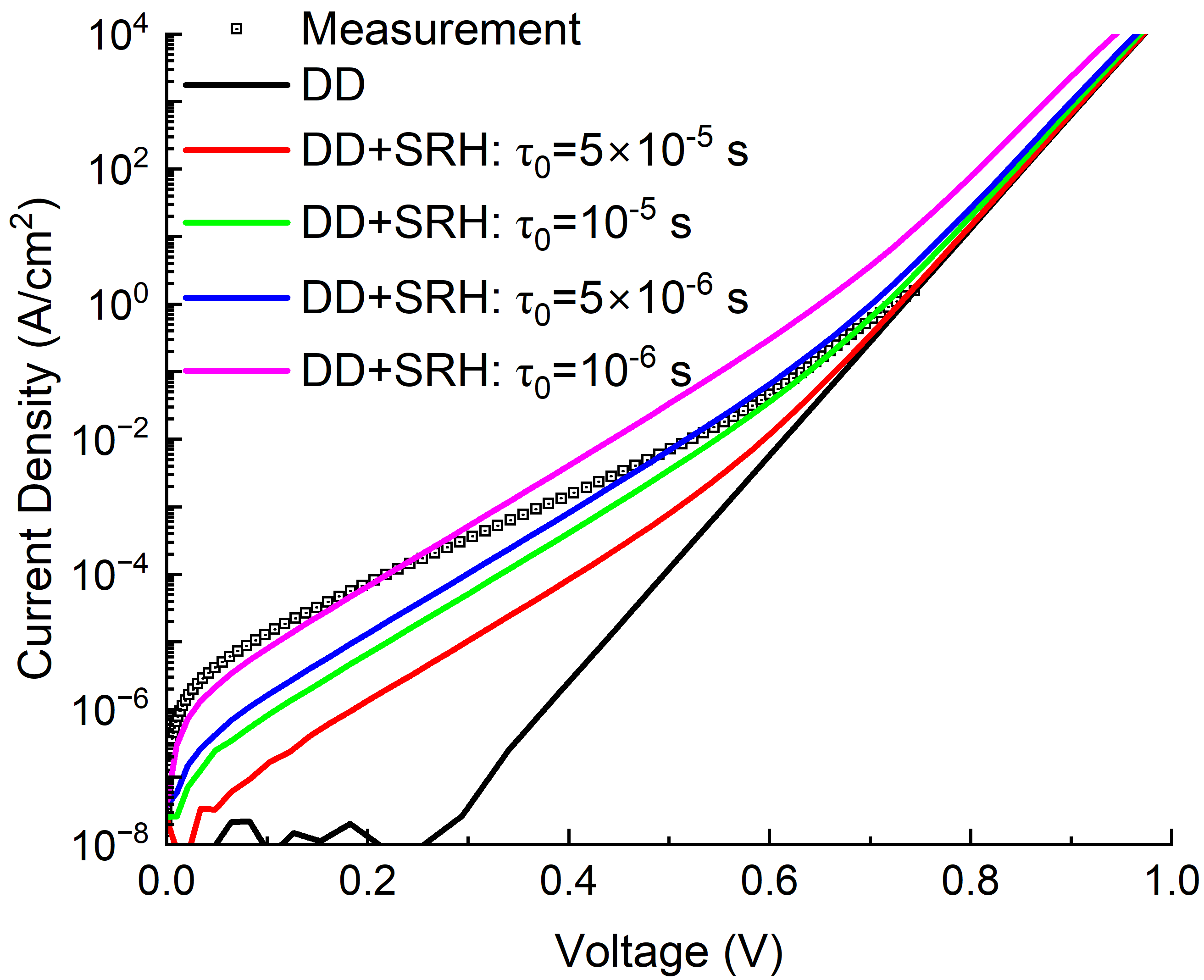}
    \caption{Analysis of the SRH contribution in the forward-bias regime at T=300~K for different filed-independent lifetime values. Note that the SRH simulations take into account the effective doping-dependent lifetimes as given in Eq.~(\ref{eq:srh-lifetime}).}
    \label{fig:SRH}
\end{figure}

Fig.~\ref{fig:SRH} shows the analysis of the temperature- and doping-independent lifetime, $\tau_0$. Here $\tau_{n0} \approx \tau_{p0} = \tau_0$ is assumed as in the TAT case for simplicity. The simulations, however, account for the effective SRH lifetime by including the effects of doping concentration and temperature using the relation defined in Appendix~\ref{Appendix:SRH}. In contrast to the TAT model, the effective SRH lifetime does not include a tunneling enhancement factor and therefore depends primarily on doping concentration and temperature rather than directly on the electric field.

As observed in Fig.~\ref{fig:SRH}, reducing the lifetime primarily increases the SRH current magnitude while preserving an approximately constant semi-logarithmic current-voltage slope across the entire bias range. 
Similar to the previous discussion for TAT, this behavior can be understood by noting that the SRH recombination rate is approximately inversely proportional to the carrier lifetime, i.e., $R_{SRH}(\tau,V) \approx (\frac{1}{\tau_0}) F(V)$, where $F(V)$ contains the voltage dependence through the carrier concentrations. Thus, reducing the lifetime mainly increases the SRH recombination rate, producing an approximately vertical shift in the semi-logarithmic I-V curve while leaving the slope unchanged. From these results, one can note that the simulations can reproduce well the measured current slope at moderate to high forward biases, i.e., between 0.5~V and 0.8~V, where TAT no longer dominates the transport.

Previous studies on silicon p$^+$-i-n$^+$ and Esaki-type junctions have reported SRH lifetimes in the sub-nanosecond to nanosecond range, with shorter lifetimes associated with higher defect densities and low-temperature-grown materials.\cite{Jorke:1993} These reported values are consistent with the temperature- and doping-independet SRH lifetime obtained in Fig.~\ref{fig:SRH}. In particular, for doping concentrations on the order of 10$^{20}$~cm$^{-3}$, field-independent lifetimes $\tau_0$ between 10$^{-5}$~s and 5$\times$10$^{-6}$~s lead to effective lifetimes on the order of nanoseconds.

\subsection{Comparison against experimental data}\label{subsec:comparison}

\begin{figure}[t]
    \centering
    \includegraphics[width=1\linewidth]{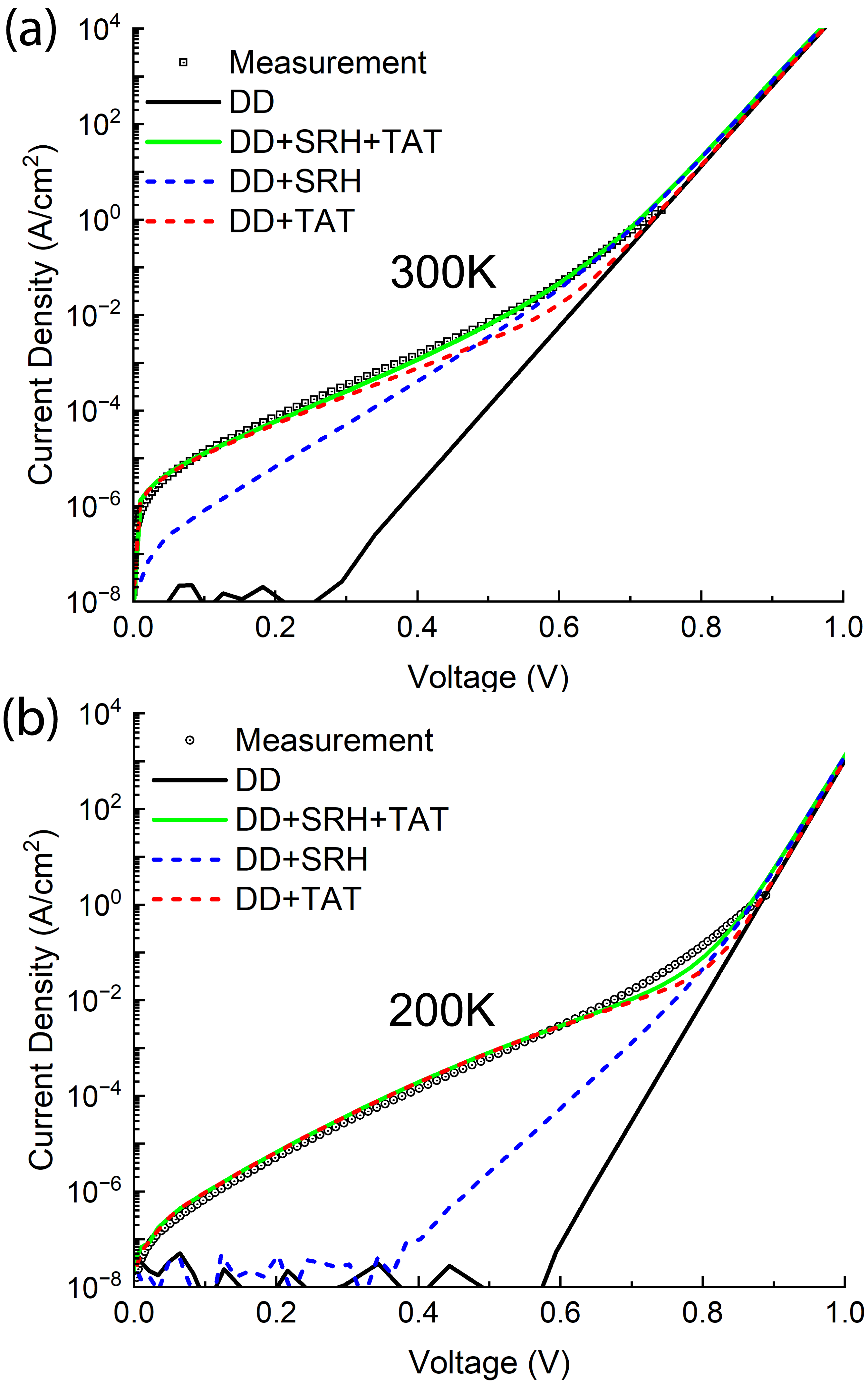}
    \caption{Simulated and measured device characteristics under forward bias at (a) T=300~K and (b) T=200~K. The contribution of each individual transport mechanism is also shown. For SRH: $\tau_0=10^{-5}$~s; for TAT: E$_t=0.5$~eV, and $\tau_0=10^{-7}$~s.}
    \label{fig:forward_bias}
\end{figure}

By comparing the transport-mechanism analyses presented in Figs.~\ref{fig:TAT} and \ref{fig:SRH} with the experimental data at 300~K in Fig.~\ref{fig:Measured-IV-with-eta}, we find that TAT dominates the low- to moderate-forward-bias regime, whereas SRH becomes increasingly important and eventually dominates at higher forward biases. The model parameter set that provides the best agreement with the measurement at 300~K consists of $\tau_0=10^{-5}$~s for SRH, and $E_t=0.5$~eV and $\tau_0=10^{-7}$~s for TAT, as shown in Fig.~\ref{fig:forward_bias}(a). Fig.~\ref{fig:forward_bias}(b) presents the simulated device response at 200~K using the same parameter set while accounting for the temperature dependence incorporated into the effective SRH and TAT lifetimes. 

Overall, the simulations in Fig.~\ref{fig:forward_bias} reproduce the measured characteristics very well across the temperature range considered and reveal a transition from TAT-dominated transport at low and moderate forward biases to SRH-dominated transport at higher biases. The origin of this transition is that, although both SRH and TAT lifetimes depend on trap density and temperature, only the effective TAT lifetime [see Eq.~(\ref{eq:tat-eq-3})] includes an explicit electric-field dependence through the tunneling enhancement factors while SRH lifetime includes a dependence on the doping density (Eq.~(\ref{eq:srh-lifetime}) in Appendix~\ref{Appendix:SRH}).

\begin{figure}[t]
    \centering
    \includegraphics[width=1\linewidth]{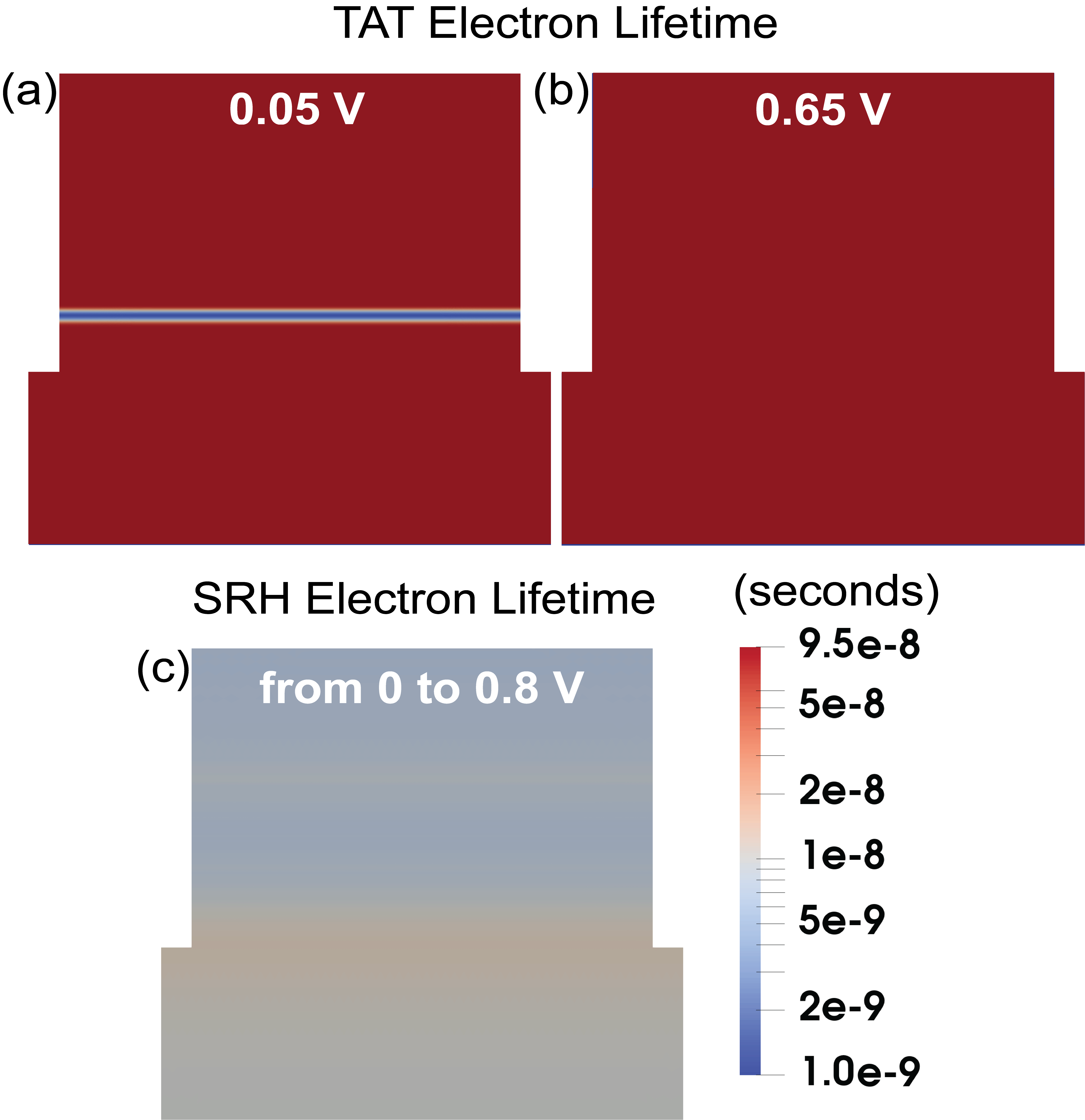}
    \caption{Analysis of the electron lifetimes at 300~K for TAT and SRH at two forward biases corresponding to the two distinct regimes where the built-in electric field across the junction is strong and weak, respectively.}
    \label{fig:foward_bias_lifetime}
\end{figure}

At low forward biases, the strong junction electric field substantially reduces the effective TAT lifetime through the tunneling enhancement factor, making TAT the dominant transport mechanism. As shown in Fig.~\ref{fig:foward_bias_lifetime}, the effective TAT lifetime under strong electric-field conditions reaches values on the order of $1.3 \times 10^{-9}$~s, whereas the effective SRH lifetime is approximately in the range of $5\times10^{-9}$~s to $9 \times10^{-9}$~s. As the forward bias increases, the electric field across the junction decreases, weakening the tunneling enhancement and increasing the effective TAT lifetime. Consequently, the relative contribution of TAT decreases, allowing SRH recombination to become the dominant transport mechanism at sufficiently high forward biases. In particular, at a voltage of 0.65~V (Fig.~\ref{fig:foward_bias_lifetime}), the effective TAT lifetime increases to approximately $9.5 \times 10^{-8}$~s, whereas the effective SRH lifetime remains on the order of $5\times10^{-9}$ s to $9 \times10^{-9}$~s. One can note that the effective SRH lifetime remains nearly constant because, unlike the TAT lifetime, it is independent of the electric field.

As the temperature decreases, the electric-field dependence of the TAT lifetime becomes increasingly more important relative to its temperature dependence. As a result, the effective TAT lifetime remains shorter than the SRH lifetime over a wider forward-bias range, extending the TAT-dominated regime while progressively reducing the relative contribution from SRH.
These trends are consistent with previous studies showing that TAT is enhanced under strong electric fields, whereas SRH recombination becomes more significant under weaker electric fields and at higher temperatures.\cite{Hurkx:1989,Hurkx:1992,Jorke:1993}

Note that the extracted parameter values should be regarded as physically reasonable estimates obtained by calibrating the simulation results to the experimental data. These values are subject to some uncertainties because nearby parameter values may also provide good agreement with the measurements. A rigorous quantification of these uncertainties would require a dedicated statistical analysis over a sufficiently broad parameter space and is beyond the scope of the present work. The primary objective of the present calibration is to identify the dominant transport mechanisms and their relative contributions across different bias and temperature regimes.

\section{Transport Mechanism for Reverse Bias}\label{subsection: Reverse response}

\begin{figure}[ht!]
    \centering
    \includegraphics[width=1.0\linewidth]{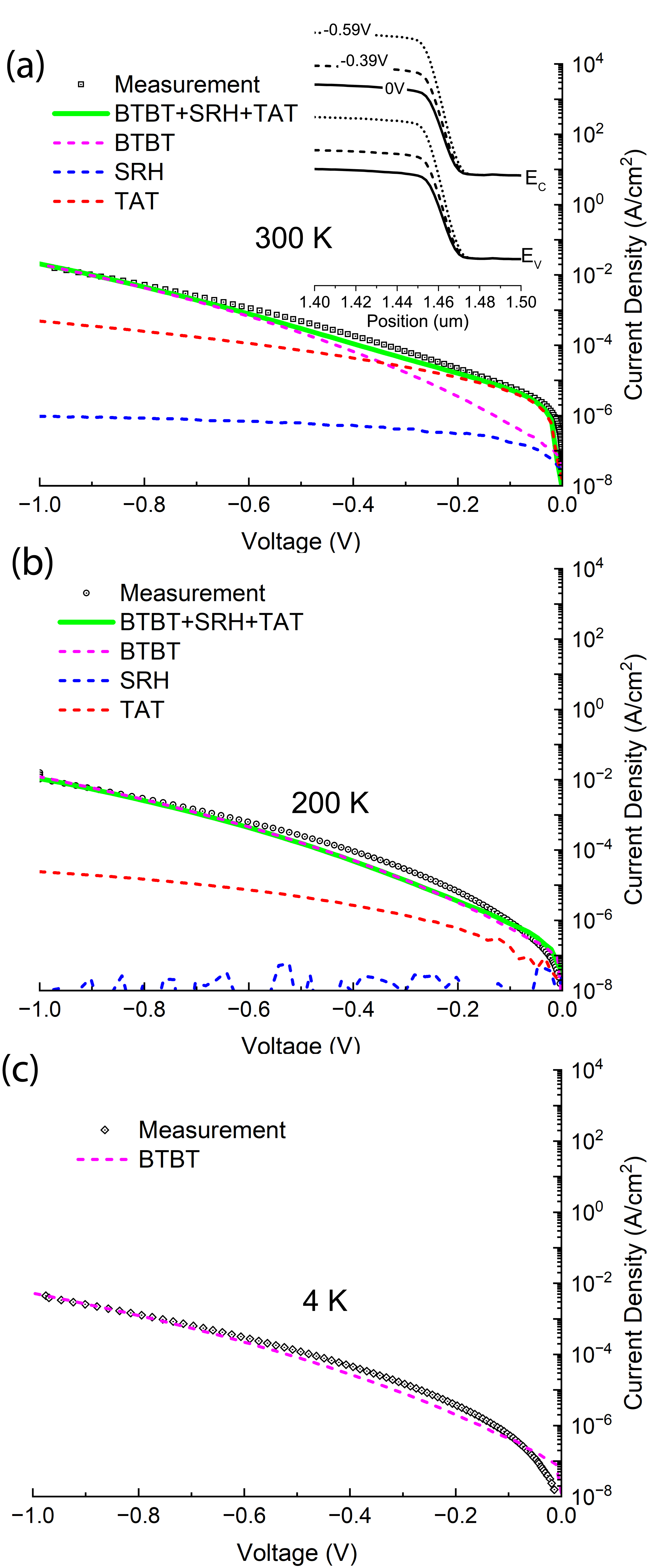}
    \caption{Simulated and measured device characteristics in reverse-bias characteristics at (a) T=300~K, (b) T=200~K, and (c) T=4~K. Inset in (a): Band structure across the junction under equilibrium and a voltage of -0.39~V and -0.59~V. 
    }
    \label{fig:reverse_bias}
\end{figure}

\begin{figure}[ht!]
    \centering
    \includegraphics[width=1\linewidth]{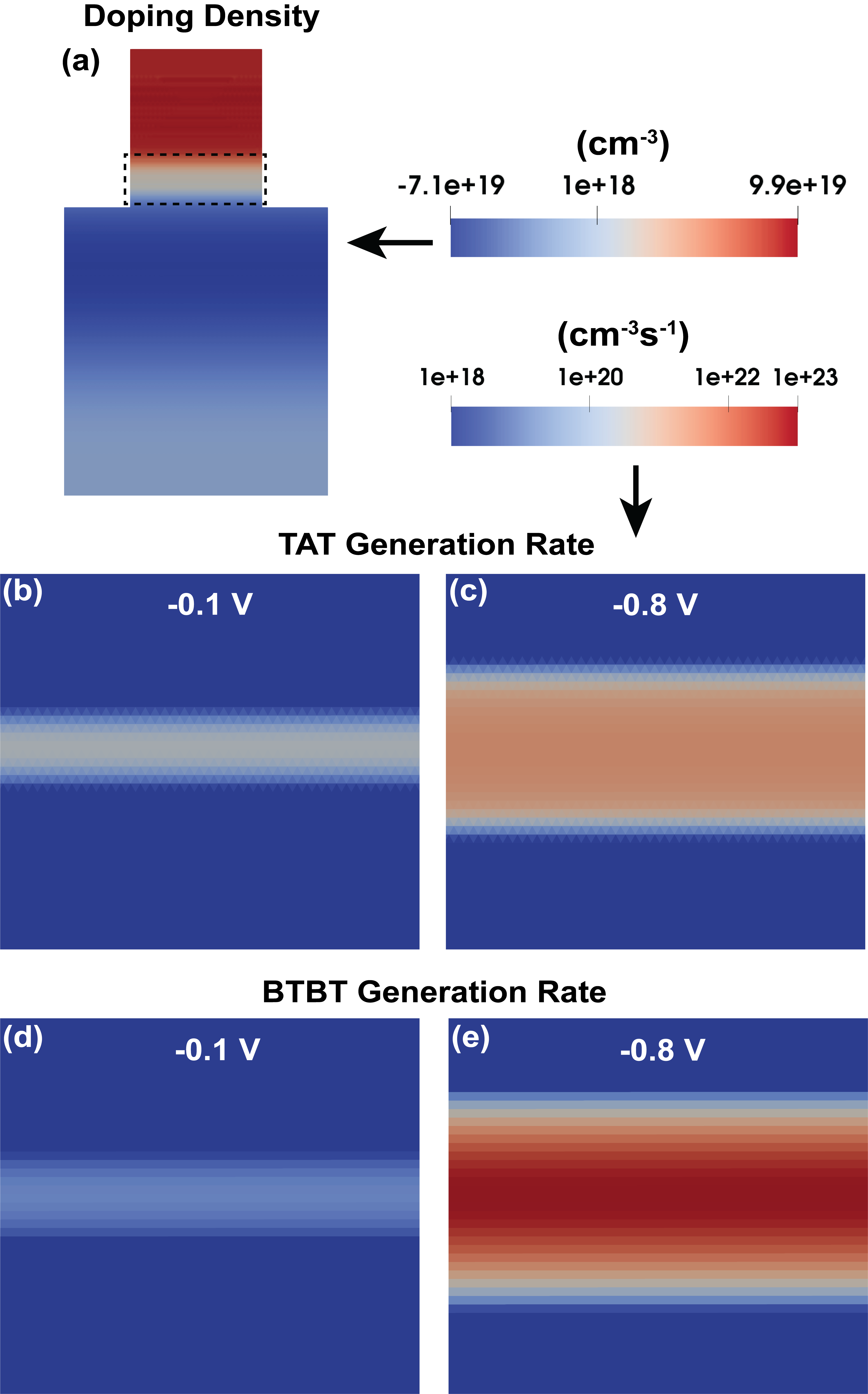}
    \caption{Analysis of TAT and BTBT generation rates at 300~K under reverse biases within the junction region marked in (a) where negative doping concentrations are p-type and positive doping concentrations are n-type. (b,c) TAT generation rate at two different negative biases. (d,e) BTBT generation rate at two different negative biases.}    \label{fig:Reverse_bias_recombination_rate_analysis}
\end{figure}

Having identified the dominant transport mechanisms for the forward-bias regime and calibrated the TAT and SRH model parameters from the forward-bias characteristics, we next investigate whether the same mechanisms and parameter set can reproduce the measured device response under reverse-bias conditions. In these reverse-bias simulations, the SRH and TAT model parameters are identical to those determined from the forward-bias analysis. Fig.~\ref{fig:reverse_bias} compares the simulated and measured reverse-bias characteristics at different temperatures together with the individual contributions from TAT, SRH, and  BTBT. The BTBT parameters used in theses simulations are listed in Appendix~\ref{Appendix:BTBT} and are consistent with the silicon parameters reported in Ref.~\onlinecite{Kao:2012}. Overall, the simulations reproduce the measured I-V characteristics both quantitatively and qualitatively over the bias range considered in the experiments.

In contrast to the forward-bias regime, the simulations in Fig.~\ref{fig:reverse_bias} show that BTBT becomes the dominant transport mechanism under reverse bias, particularly at low temperatures. This behavior is expected because increasing the reverse bias simultaneously enhances both the electric field across the junction and the overlap between occupied valence-band states on the p-type side and empty conduction-band states on the n-type side, whereas reduces the tunneling distances, as illustrated in the inset of Fig.~\ref{fig:reverse_bias}(a). Together, these effects substantially increase the interband tunneling probability. Physically, this transport mechanism corresponds to interband tunneling, where valence electrons on the p-type side tunnel into the conduction band on the n-type side. This transport mechanism is commonly referred to as Zener tunneling and is responsible for the monotonic increase in the reverse current observed in previous measurements for silicon Esaki diodes.\cite{Oehme:2009,Oehme:2010,Oehme:2012}

While BTBT dominates at low temperatures, the simulations further reveal that TAT contributes significantly at elevated temperatures and low reverse biases. In particular, at 300~K, TAT clearly dominates the transport between 0~V and approximately $-0.35$~V, whereas BTBT becomes dominant at larger reverse biases. Fig.~\ref{fig:Reverse_bias_recombination_rate_analysis} further illustrates this transition by comparing the spatial TAT and BTBT generation rates at $-0.1$~V and $-0.8$~V. 
A similar qualitative transition has been reported previously for silicon tunnel junctions operating above room temperature in Refs.~\onlinecite{Hurkx:1989,Hurkx:1992}. As the temperature decreases to 200~K and 4~K (Fig.~\ref{fig:reverse_bias}(b) and (c), respectively), the TAT contribution rapidly diminishes and BTBT becomes the dominant transport mechanism over the entire reverse-bias range investigated, with TAT contributing only marginally to the total current. The simulations in Fig.~\ref{fig:reverse_bias} also indicate that the SRH mechanism provides a negligible contribution throughout the reverse-bias range investigated compared to BTBT and TAT, especially at low temperatures.

An important observation from Figs.~\ref{fig:Measured-IV-with-eta}, \ref{fig:forward_bias} and \ref{fig:reverse_bias} is that the BTBT contribution exhibits a much weaker temperature dependence than the TAT and SRH contributions. This behavior in consistent with the fact that BTBT is a quantum-mechanical tunneling process governed primarily by the junction electric field and the overlap between occupied and empty electronic states, without the involvement of intermediate defect states. Consequently, the BTBT probability depends only weakly on temperature. In contrast, TAT and SRH rely on carrier capture and emission through localized defect states, which are thermally activated processes and therefore exhibit a much stronger temperature dependence. As the temperature decreases, the TAT and SRH contributions are progressively suppressed, whereas BTBT remains largely unaffected.

The relative contribution of TAT is expected to increase further at temperatures above room temperature. Higher temperatures enhance carrier capture and emission through localized defect states, increasing the TAT current particularly at low reverse biases where the electric field is insufficient for BTBT to dominate. Consequently, the crossover voltage between TAT- and BTBT-dominated transport is expected to shift toward larger reverse biases, in agreement with previous observations reported for silicon tunnel junctions operating above room temperature.\cite{Hurkx:1989,Hurkx:1992}

\subsection{Avalanche}\label{subsection:Avalanche}

\begin{figure}[t]
    \centering
    \includegraphics[width=1\linewidth]{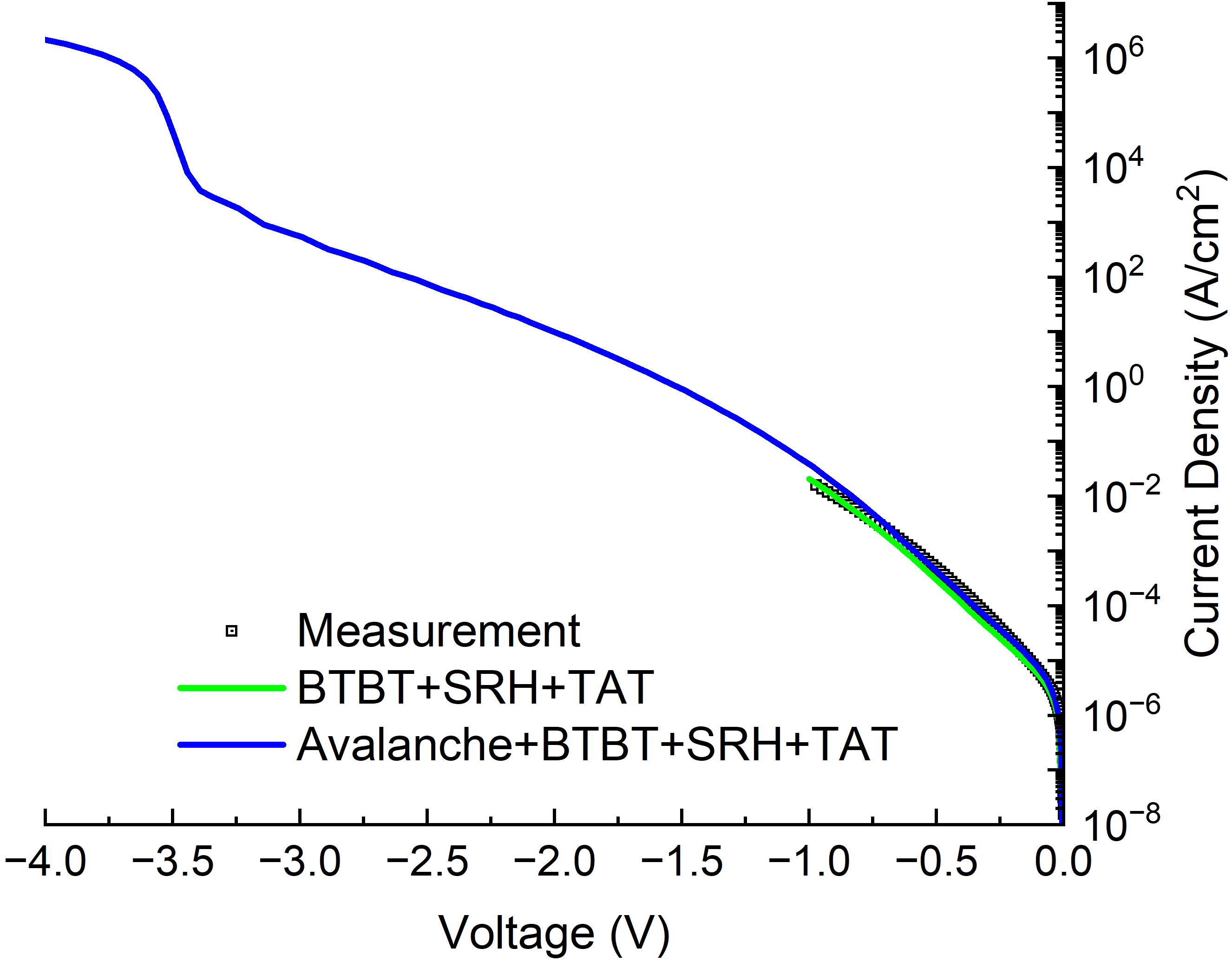}
    \caption{Effect of avalanche generation on the device under reverse biases at T=300~K.}
    \label{fig:avalanche}
\end{figure}

Under reverse-bias conditions, avalanche generation can become relevant when the electric field across the junction is sufficiently strong to accelerate carriers to energies capable of generating additional electron-hole pairs through impact ionization. This process leads to carrier multiplication and a rapid increase in the reverse current. Fig.~\ref{fig:avalanche} shows the contribution of avalanche generation under reverse bias at 300~K. The simulation indicates that avalanche generation provides a negligible contribution over the bias range investigated (between 0~V and 1~V), confirming that the measured reverse current is primarily governed by BTBT and TAT mechanisms. This behavior is expected because the electric field remains below the threshold required for significant impact ionization throughout the experimental bias range. Furthermore, the simulation predicts that avalanche breakdown occurs only at substantially larger reverse biases, approximately 3.5~V.

\section{Temperature-Dependent Transport Analysis}\label{sec:temp-dep-transport}

In this section, we investigate how the device response and the interplay among the transport mechanisms evolve over a wide temperature range. 

\begin{figure}[t]
    \centering
    \includegraphics[width=1.0\linewidth]{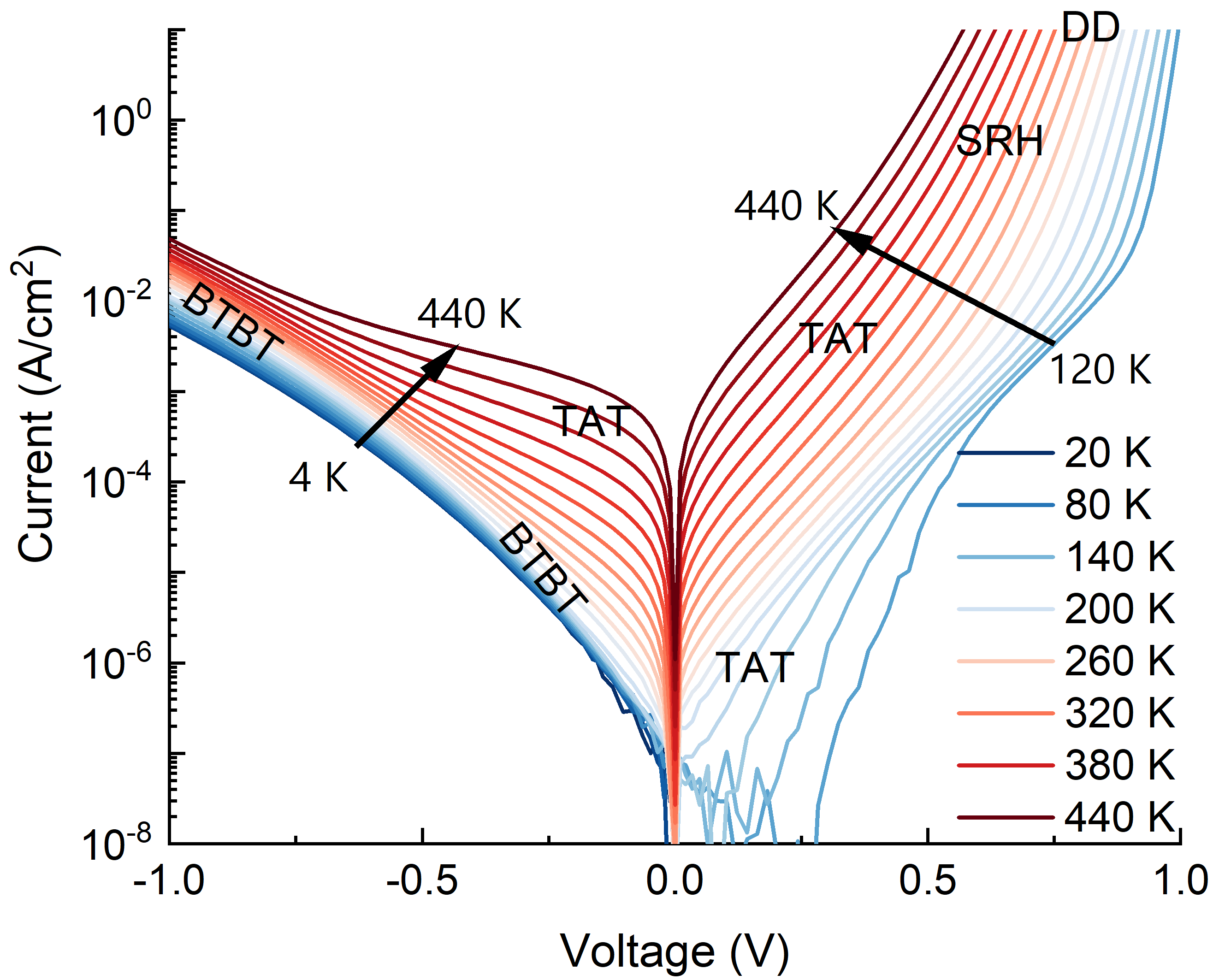}
    \caption{Simulated I-V characteristics over a temperature range of $4$ to $440$ K, including the contributions of BTBT, TAT, and SRH transport mechanisms. The labels in the figure represent the primary transport mechanism.}
    \label{fig:I-V_temp_study}
\end{figure}

Fig.~\ref{fig:I-V_temp_study} shows the simulated device characteristics under both forward and reverse biases across temperatures ranging from 4~K to 440~K. The simulations include all relevant transport mechanisms in the device from the above studies, namely SRH, TAT, and BTBT, employing the above fitted parameters. Overall, the results reproduce the key feature observed in the measured I-V characteristics shown in Fig.~\ref{fig:Measured-IV-with-eta} and in Ref.~\onlinecite{Hong:2026}. 
Under forward biases, the current exhibits a strong temperature dependence while maintaining a nearly constant slope over a range of voltages and temperatures, with TAT dominating at low-to-moderate biases, followed by increasing SRH contribution at higher biases and DD at even higher biases. For lower temperatures, DD comes into play at somewhat lower forward bias.
Under reverse bias, a pronounced temperature dependence is observed at low biases and high temperatures, appearing as a characteristic bump in the I-V curves. This behavior indicates that TAT mechanisms dominate in this regime. In contrast, at higher reverse biases or lower temperatures, the temperature dependence diminishes, indicating a transition to BTBT-dominated transport. In addition, below approximately 20~K, the simulated reverse-bias characteristics become nearly temperature independent.

\section{Conclusions} \label{sec: Conclusions}

We have developed and validated an extended Schenk trap-assisted tunneling (TAT) model by incorporating Fermi-Dirac statistics, enabling the modeling of carrier transport in heavily doped silicon tunneling junctions over a broad range of temperature and voltage conditions. The extended model enables physically consistent TAT simulations at temperatures below which the original Maxwell-Boltzmann (MB)-based formulation remains inadequate. This extension therefore provides more flexible TCAD modeling capability for investigating tunneling devices over a wider range of temperatures.

Application of the enhanced TAT model and extensive TCAD simulations over a broad temperature and voltage range reveal distinct bias- and temperature-dependent transport regimes in the measured silicon tunneling device. Under forward biases, band-to-band tunneling (BTBT) is negligible, while TAT dominates at low to moderate biases and Shockley-Read-Hall (SRH) mechanism becomes increasingly important at higher biases and elevated temperatures. Under reverse bias, BTBT dominates at low temperatures and high reverse biases, whereas TAT becomes important at low reverse biases and high temperatures. The successful capture of these transitions using the same set of calibrated parameters demonstrates the effectiveness of the modeling framework to resolve competing transport mechanisms across diverse operating conditions.

Finally, the absence of negative differential conductance (NDC) and the negligible BTBT contribution under forward bias suggest that the doping profile of the present CVD-fabricated device may not be sufficiently abrupt to provide strong forward-bias interband tunneling. This finding highlights the importance of fabrication approaches capable of producing sharper doping profiles, such as molecular beam epitaxy (MBE) or atomically precise advanced manufacturing (APAM),\cite{Ward:2020,Schofield:2025,Anderson:2025,Gao:2021.2,Lu:2021} where narrower junction transition regions can increase the electric field, reduce the tunneling distance, and enhance BTBT. Such improved junction abruptness may enable Esaki-diode-like NDC behavior and potentially reveal additional tunneling-driven transport physics not accessible in the present device.

\begin{acknowledgments}
We thank Andrew J. Leenheer for helpful conversations.
This work was supported by the U.S. Department of Energy (DOE) Advanced Materials and Manufacturing Technologies Office (AMMTO) project BEATS. 
Sandia National Laboratories is a multimission laboratory managed and operated by National Technology and Engineering Solutions of Sandia, LLC., a wholly owned subsidiary of Honeywell International, Inc., for the U.S. Department of Energy’s National Nuclear Security Administration under contract DE-NA-0003525. Work was performed, in part, at the Center for Integrated Nanotechnologies, a U.S. DOE, Office of Basic Energy Sciences user facility. This paper describes objective technical results and analysis. Any subjective views or opinions that might be expressed in the paper do not necessarily represent the views of the U.S. Department of Energy or the United States Government. This article has been authored by an employee of National Technology \& Engineering Solutions of Sandia, LLC under Contract No. DE-NA0003525 with the U.S. Department of Energy (DOE). The employee owns all right, title and interest in and to the article and is solely responsible for its contents. The United States Government retains and the publisher, by accepting the article for publication, acknowledges that the United States Government retains a non-exclusive, paid-up, irrevocable, world-wide license to publish or reproduce the published form of this article or allow others to do so, for United States Government purposes. The DOE will provide public access to these results of federally sponsored research in accordance with the DOE Public Access Plan https://www.energy.gov/downloads/doe-public-access-plan. 
\end{acknowledgments}

\section*{AUTHOR DECLARATIONS}

\subsection*{Conflict of Interest} 
The authors have no conflicts to disclose.

\section*{DATA AVAILABILITY}
The data that support the findings of this study are available from the corresponding author upon reasonable request

\appendix


\include{transport_models}

\bibliography{main}

\end{document}

%% file: transport_models.tex
\section{Band-to-Band Tunneling Model}\label{Appendix:BTBT}

Band-to-band tunneling (BTBT) is modeled as a field-dependent generation rate which is given by \cite{Kane:1960, Wong:2020}
\begin{equation}
G_{BTBT} = \left( \frac{|F-F_{eq}|}{F_{eq}} \right)^\beta \alpha A \left( \frac{F}{F_0} \right)^\gamma \textrm{exp} \left(-\frac{B}{F} \right). 
\label{eq:modified-b2bt}
\end{equation}
The value of $\alpha$ can be 0, -1, or 1. When $\alpha = -1$, $G_{BTBT} < 0$, which indicates the BTBT is a pure recombination process. When $\alpha = 1$, $G_{BTBT} > 0$, which indicates the BTBT is a pure generation process. $F_0$ = 1 V/cm, $F$ is the electric field magnitude in V/cm unit, and $F_{eq}$ is the simulated, position-dependent electric field (magnitude) in a device at a zero-current condition. $\beta=0$ is used in this work.

$A$, $B$, and $\gamma$ are fitting parameters that can be computed for Si, for both direct and indirect BTBT processes,  according to Ref. \cite{Kao:2012}, as follows
\begin{eqnarray}
A_{dir} &=& \left[ \frac{g \pi m_r^{1/2} \left(q F_0 \right)^2} {9 h^2 \left( E_g^\Gamma \right)^{1/2}} \right] \left( \frac{9}{\pi^2} \right), \label{eq:b2bt-Adir} \\
B_{dir} &=& \frac{\pi^2 m_r^{1/2} \left( E_g^\Gamma \right)^{3/2} } {q h}, \label{eq:b2bt-Bdir} \\
A_{ind} &=& \frac{g \left(m_c m_v \right)^{3/2} \left(1+2 N_{TA} \right) D_{TA}^2 \left( q F_0\right)^{5/2} } {2^{21/4} h^{5/2} m_r^{5/4} \rho E_{TA} E_g^{7/4}}, \label{eq:b2bt-Aind} \\
B_{ind} &=& \frac{2^{7/2} \pi m_r^{1/2} E_g^{3/2}} {3 q h}, \label{eq:b2bt-Bind}
\end{eqnarray}
where $g$ is the degeneracy factor; $m_r$ is the reduced tunneling mass; $q$ is the elementary charge; $h$ is Planck’s constant; $E_g^\Gamma$ is the direct band gap at the $\Gamma$ point; $m_v$ ($m_c$) is the valence (conduction) band density of states effective mass; $N_{TA} = 1/[\exp(E_{TA}/k_BT) - 1]$ is the occupation number of the transverse acoustic phonon at temperature $T$ with $k_B$ being the Boltzmann constant; $D_{TA}$ is the deformation potential of transverse acoustic phonons; $\rho$ is the mass density; $E_{TA}$ is the transverse acoustic phonon energy; and $E_g$ is the minimum band gap. 

For indirect transitions, the conduction and valence band density of state effective mass $m_c$ and $m_v$ are respectively given by $m_c=\left(m_l m_t^2 \right)^{1/3}$ and $m_v=\left(m_{lh}^{3/2} + m_{hh}^{3/2} \right)^{2/3}$, where $m_l$ and $m_t$ are the electron longitudinal and transverse effective masses, and $m_{lh}$ and $m_{hh}$ are the light and heavy hole effective masses, respectively. The reduced tunneling mass $m_r$ is computed as $m_r = m_e m_{lh} / (m_e + m_{lh})$ where  $m_e$ is the lightest electron effective mass along a certain tunneling direction. 

The degeneracy factor $g$ is expressed as $g = 2 \times g_v \times g_c$ where 2 is due to the electron spin degeneracy factor, the valence band valley degeneracy are $g_v=1$ since we only consider the lightest carriers, and the conduction band valley degeneracy $g_c$ depends on the tunneling direction.

The physical parameters used in computing $A_{dir}$, $B_{dir}$, $A_{ind}$, and $B_{ind}$ for Si in this work are  $\rho=2329$~kg/m$^3$, $D_{TA}=2.74 \times 10^{10}$~eV/m, $E_{TA}=19$~meV, $E_g=1.12$~eV, $E_g^\Gamma=3.4$~eV, $m_l=0.98$, $m_t=0.19$, $m_{lh}=0.16$, $m_{hh}=0.49$, $m_c=0.328$, and  $m_v=0.549$.  Note all effective masses are in the unit of free electron mass. For direct BTBT in Si, a degeneracy factor $g=2$ and a reduced tunneling mass $m_r=0.08$ are used. For indirect BTBT along the [100] direction in Si, a degeneracy factor $g=8$, an electron effective mass $m_e=m_t$, and a reduced tunneling mass $m_r=0.087$ are used.

\section{Mid-Gap SRH Recombination Model}\label{Appendix:SRH}

The standard Shockley–Read–Hall (SRH) recombination model is given by
\begin{equation}
  R_{\text{SRH}} = \frac{np - n_0 p_0}{\tau_p \left( n + \sqrt{n_0 p_0} \right) +
  \tau_n \left( p + \sqrt{n_0 p_0} \right)}.
\end{equation}
Here $n_0$ and $p_0$ are equilibrium electron and hole densities, respectively. In the case of Boltzmann statistics, the product $n_0
p_0$ is given by $n_i^2$, where $n_i$ is the intrinsic carrier concentration in a material. This model represents the effect of
mid-band gap traps.

The parameters $\tau_n$ and $\tau_p$ represent electron and hole
lifetimes, respectively, and are dependent on dopant concentration and temperature as

\begin{equation}
\label{eq:srh-lifetime}
  \tau _{n(p)}=  \frac{\tau_{n(p)0}}{1+ \frac{N_D + N_A} {N_{srh}}} \left( \frac{T}{300} \right)^\alpha.
\end{equation}\label{SRH:lifetime}
where $N_D$ and $N_A$ are the donor and acceptor concentrations, respectively, and $N_{srh}=5.0 \times 10^{16}$~$\frac{1}{\text{cm}^3}$ is a material dependent parameter and $\alpha=-1.5$.

\section{Avalanche Generation Model}\label{Appendix:Avalanche}

The generation rate of electron-hole pairs produced by the avalanche effect can described by 
\begin{eqnarray}
  \label{eq:ava_gen}
  G & = & \alpha_n(F_{n},T) \frac{|\mathbf{J}_n|}{q} + \alpha_p(F_{p},T) \frac{|\mathbf{J}_p|}{q}
\end{eqnarray}
where $\alpha_n(F_n,T)$ and $\alpha_p(F_p,T)$ are the electron and hole ionization coefficients, $F_{n}$, $F_{p}$ are electron and hole driving force magnitudes and  $\mathbf{J}_e$, $\mathbf{J}_h$ are the electron and hole current density vectors. There are different models to compute the electron and hole ionization coefficients. In this work, we have used the well-known Selberherr model \cite{Selberherr:1984,SCHUTZ:1982}, where these coefficient are computed as
\begin{eqnarray}
  \label{eq:ava_gen_selberherr1}
  \alpha_n & = & \alpha_n^\infty \left( T \right) \exp\left[ -\left(
    \frac{F_n^{crit}}{F_n} \right)^{\delta_n} \right]~\nonumber \\ 
  \alpha_p & = & \alpha_p^\infty \left( T \right) \exp\left[ -\left(
    \frac{F_p^{crit}}{F_p} \right)^{\delta_p} \right]
\end{eqnarray}
$F_n^{crit}$ and $F_p^{crit}$ are the critical fields and can be computed by 
\begin{eqnarray}
  \label{eq:ava_gen_selberherr2}
  F_n^{crit} & = & \frac{E_g\left(T\right)}{q
    \lambda_n\left(T\right)}\nonumber \\
  F_p^{crit} & = & \frac{E_g\left(T\right)}{q
    \lambda_p\left(T\right)},
\end{eqnarray}
where $\lambda_n$ and $\lambda_p$ are the optical-phonon mean free paths for electrons and holes, given by
\begin{eqnarray}
  \label{eq:ava_gen_selberherr3}
  \lambda_n \left(T\right) & = & \lambda_n^0  
        \ \tanh\left(\frac{\hbar \omega_n}{2k_BT}\right) \nonumber \\ 
  \lambda_p \left(T\right) & = & \lambda_p^0 
        \ \tanh \left(\frac{ \hbar \omega_p}{2k_BT}\right).
\end{eqnarray}
The coefficients $\alpha_n^\infty$ and $\alpha_n^\infty$ are polynomial functions of temperature
\begin{eqnarray}
  \label{eq:ava_gen_selberherr4}
  \alpha_n^\infty \left( T \right) & = & a_{0n} + a_{1n} T + a_{2n} T^2 \nonumber \\
  \alpha_p^\infty \left( T \right) & = & a_{0p} + a_{1p} T + a_{2p} T^2 
\end{eqnarray}
with the constant coefficients $a_{0n}$, $a_{1n}$, $a_{2n}$,$a_{0p}$, $a_{1p}$ and $a_{2p}$.

\section{Statistics Effect in TAT Model}\label{Appendix:Stat-Effect-in-TAT}

\begin{figure}[t]
    \centering
    \includegraphics[width=1\linewidth]{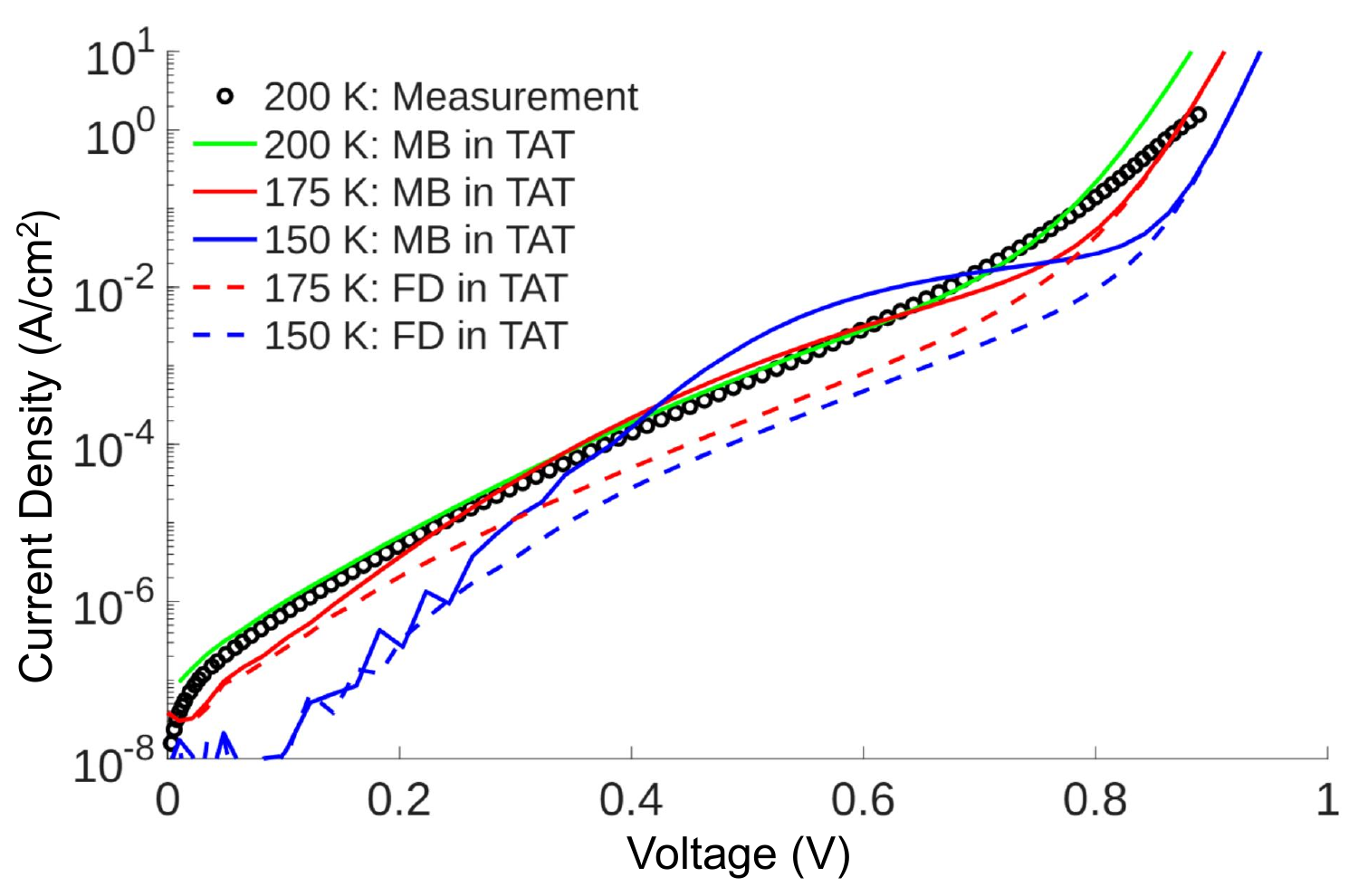}
    \caption{Comparison of the simulated forward-bias I-V characteristics obtained using MB and FD statistics in the TAT model at 200, 175, and 150~K. Clearly, the FD-based TAT model removes the nonphysical behavior produced by the MB approximation in the TAT model.}
    \label{fig:cmprsn-tat-fd-mb}
\end{figure}

In this section, we study the effects of the carrier statistics used in the Schenk TAT model on simulation results. The original Schenk TAT model uses the Maxwell-Boltzmann (MB) statistics as discussed in Sec.~\ref{sec: extended TAT}. From Fig.~\ref{fig:cmprsn-tat-fd-mb}, we see that the MB-based TAT model reproduces the measured currents very well at 200 K by comparing the green solid curve with the measured data. However, for temperatures below 200 K, e.g., 175 K and 150 K, the MB-based Schenk TAT model produces nonphysical results, which is evidenced by the fact that simulated currents (e.g., blue curve) at a lower temperature (e.g., 150 K) are larger than those at a higher temperature (e.g., 200 K) for the voltage range of approximately 0.4-0.7~V. These abnormal higher currents at a lower temperature were not observed in experimental data. We speculated that the abnormal current behavior is due to the use of MB statistics in the TAT model, which is supported by the observation that the behavior becomes worse as the temperature decreases from 175 K to 150 K. In Sec.~\ref{sec: extended TAT}, the inclusion of the Fermi-Dirac (FD) statistics in the Schenk TAT model is discussed in detail. The simulated currents using the proposed FD-based TAT model are shown as the dashed curves in Fig.~\ref{fig:cmprsn-tat-fd-mb}. Clearly, the non-physical behavior is removed and physically consistent simulation results are obtained. Specifically, as shown by the dashed curves, the simulated current at any given voltage decreases with decreasing temperature when the FD-based TAT model is used, as expected.  

